\documentclass[10pt,journal]{IEEEtran}
\usepackage{amssymb}

\usepackage{slashbox}

\usepackage{ifpdf}

 \ifpdf

 \else

 \fi

\ifCLASSINFOpdf
   \usepackage[pdftex]{graphicx}

   \graphicspath{{../pdf/}{../jpeg/}}

   \DeclareGraphicsExtensions{.pdf,.jpeg,.png}
\else

   \usepackage[dvips]{graphicx}

   \graphicspath{{../eps/}}

   \DeclareGraphicsExtensions{.eps}
\fi

\usepackage[cmex10]{amsmath}

\usepackage{algorithmic}

\usepackage{array,cite}

\usepackage{mdwmath}
\usepackage{mdwtab}

\usepackage{eqparbox}

\usepackage[tight,footnotesize]{subfigure}

\usepackage[caption=1]{caption}

\usepackage[caption=false,font=footnotesize]{subfig}

\usepackage{stfloats}

\fnbelowfloat

\usepackage{url}

\allowdisplaybreaks[4]

\begin{document}

\title{\mbox{}\\
\textsc{On the Capacity Region of Cognitive Multiple Access over White Space Channels}}

\author{\normalsize \authorblockN{Huazi~Zhang, Zhaoyang~Zhang,~\IEEEmembership{Member,~IEEE}, and Huaiyu~Dai,~\IEEEmembership{Senior Member,~IEEE}}
\thanks{Manuscript received Apr. 18, 2012; revised Sep. 3, 2012.}
\thanks{
Huazi~Zhang (E-mail: {\tt thomas25@163.com}) and Zhaoyang~Zhang (E-mail: {\tt ning\_ming@zju.edu.cn}, Phone \& Fax: +86-571-87952060) are with the Department of Information Science and Electronic Engineering, Zhejiang University, Hangzhou, China. Huaiyu~Dai (E-mail: {\tt hdai@ncsu.edu}, Phone: (919) 513-0299, Fax: (919) 515-5523) is with the Department of Electrical and Computer Engineering, NC State University, Raleigh, NC 27695.
}
\thanks{This work was supported in part by the National Key Basic Research Program of China (Nos. 2009CB320405 and 2012CB316104), the National Natural Science Foundation of China (No.60972057), the Zhejiang Provincial Natural Science Foundation of China (No.LR12F01002), the Supporting Program for New Century Excellent Talents in University (No. NCET-09-0701), and the US National Science Foundation under Grants CCF-0830462 and ECCS-1002258.}}

\markboth{IEEE Journal on Selected Areas in Communications,~Vol.~31, No.~11, Nov.~2013}%
{Shell \MakeLowercase{\textit{et al.}}: Bare Demo of IEEEtran.cls
for Journals}

\maketitle

\begin{abstract}
Opportunistically sharing the white spaces, or the temporarily unoccupied spectrum licensed to the primary user (PU), is a practical way to improve the spectrum utilization. In this paper, we consider the fundamental problem of rate regions achievable for multiple secondary users (SUs) which send their information to a common receiver over such a white space channel. In particular, the PU activities are treated as on/off side information, which can be obtained causally or non-causally by the SUs. The system is then modeled as a multi-switch channel and its achievable rate regions are characterized in some scenarios. Explicit forms of outer and inner bounds of the rate regions are derived by assuming additional side information, and they are shown to be tight in some special cases. An optimal rate and power allocation scheme that maximizes the sum rate is also proposed. The numerical results reveal the impacts of side information, channel correlation and PU activity on the achievable rates, and also verify the effectiveness of our rate and power allocation scheme. Our work may shed some light on the fundamental limit and design tradeoffs in practical cognitive radio systems.

\end{abstract}


\IEEEpeerreviewmaketitle

%\newpage
%\setcounter{page}{1}

\section{Introduction}

\subsection{Motivation}
To solve the dilemma between the ever increasing bandwidth demand and the actual under-utilization of spectrum resource
\cite{NSFreport}, FCC has allowed unlicensed devices to opportunistically access the temporarily unoccupied TV spectrum, namely \emph{white spaces}\cite{FCCorder}. To this end, Cognitive Radio (CR) techniques that adopt the ``sense and access" paradigm, in which the Secondary Users (SUs) identify the activity of the Primary User (PU) before accessing the \emph{white space channels}, has been extensively studied in the past years.

Recently, more and more effort has been paid to apply CR into practical systems. The above idea of opportunistically accessing the white spaces is one of the most practical ways for CR applications, which has been implemented in several systems such as the Cognitive Radio Sensor Networks (CRSN) \cite{CRSN,Cluster-CRSN,Zhang-2011ICC-JSCS,Zhang-2011INFOCOM-CS} and IEEE 802.22 Wireless Regional Area Networks (WRAN) \cite{802.22}. In both networks, SUs are grouped into clusters according to their locations and channel separation \cite{Cluster-CRSN}. The cluster members (CMs) transmit to the cluster head (CH) through a common white space channel in the uplink, which forms a special cognitive multiple access channel (CogMAC) that has the white space as the common media.

In such a special CogMAC, a fundamental problem exists, that is, what's the fundamental limit when multiple secondary users are supposed to send their information to the CH over a common white space? The answer will allow us to better understand the performance of practical cognitive radio systems. Equipped with the spectrum sensing capability, SUs can obtain the states of PU activities, either causally or non-causally. Thus, PU activities can be viewed as side information about the channel state at both cognitive transmitters and receivers, and information theoretic results on channel capacity with side information \cite{MAC-side} may be explored to reveal the fundamental limit and various tradeoffs.

\subsection{Related Works}
The existing studies on the capacity of cognitive channels are
mainly conducted in the context of interference channels \cite{Gridlock-CR-IT}, further
divided into the underlay \cite{Underlay} and overlay
\cite{Overlay1,Overlay2} cases. The former assumes that
the SU has the channel knowledge and can control
its transmission power to restrict the interference to the PUs. In contrast, the overlay approach models the coexistent
communications as the ``interference channel with degraded message
sets (IC-DMS)" \cite{Overlay3-ICDMS,Overlay4-ICDMS}, and employs intricate coding schemes for capacity enhancement.

The capacity of the cognitive MAC fading channel is studied recently in \cite{fading-C-MAC}, in which the ergodic sum-rate is obtained under the interference-power constraint. In \cite{MID} the impact of multiuser interference diversity is further exploited on the
the capacity regions of various CR networks, including the cognitive MAC channel.

Recent works on the throughput of cognitive networks are fueled by the seminal work \cite{NetworkCapacity}. In particular, it has been shown that both the primary and secondary network can achieve the throughput scaling as two standalone networks \cite{NetworkCapacityCR,NetworkCapacityCR2,NetworkCapacityCR_CLi,Sun-Achievable-RelayCRN} under various situations.

In the above study, concurrent transmission of the primary and
secondary systems is allowed. To effectively control the
interference, either channel gains of the SU-PU links or PU messages are assumed known at the secondary transmitter. However, many
practical cognitive systems only determine the existence of PU transmissions through spectrum sensing, and access the \emph{white space channels} to avoid undesired interference. In these applications, locally sensed PU activities become the major factors influencing the rate regions of SUs.

\subsection{Summary of Contributions}
To the best of our knowledge, the pioneering work \cite{Two-switch}
is the first to consider the ``sense and access" paradigm, in which
the PU activities sensed at the cognitive transmitter and receiver
are modeled as on/off side information, and the capacity of a
two-switch cognitive channel is explored. Motivated by this work, in this paper we extend the study to a cognitive multiple access
scenario. The contributions of this
paper are summarized below:

\begin{enumerate}
\item{Achievable Rate Regions of the Cognitive MAC Channel:}
By viewing the primary user activities around the transmitters and receiver as side information, we model the memoryless cognitive MAC channel as a three-switch MAC channel. The achievable regions of the three-switch MAC channel are derived for two scenarios with independent causal and non-causal side information at the transmitters, and a special case in which the receiver has strong spectrum sensing capacity. The insights gained from our results may shed some light on the design of practical systems.

\item{Outer and Inner Bounds:}
The capacity regions of the three-switch MAC channel are defined implicitly in the form of mutual information. We further obtain explicit outer and inner bounds of the capacity region by assuming additional side information at the transmitters or receiver. It is found that the outer and inner bounds coincide in two special cases: when the side information between the transmitters and receiver are highly correlated, or when the states of PU signals change slowly.

\item{Sum-rate Optimal Rate/Power Allocation:}
We optimize the rate and power allocation between transmitters when both the transmitters and receiver have global side information, and analyze the impact of two parameters, PU occupation probability and correlation in side information, on the sum rate. The results may serve as a guideline for the design of power/rate allocation algorithm in practical sensing-based CR uplink transmission systems.

\item{Extension to the fading scenario:}
We further analyze a general model with fading and interference, in which the receiver is active all the times. The results provide a guideline for the power/rate allocation algorithm design under channel fading and interference.
\end{enumerate}

The rest of the paper is organized as follows. In Section II, we
introduce the memoryless cognitive MAC channel and model it as a
three-switch MAC channel. The main results of rate expressions,
bounds and rate/power allocation are listed in Section III, with
relevant derivations and analysis given in Section IV. Numerical results are presented in Section V, and the whole paper is
concluded in Section VI.

\section{Problem Formulation and System Model}
\begin{figure}[t] \centering
\includegraphics[width=0.4\textwidth]{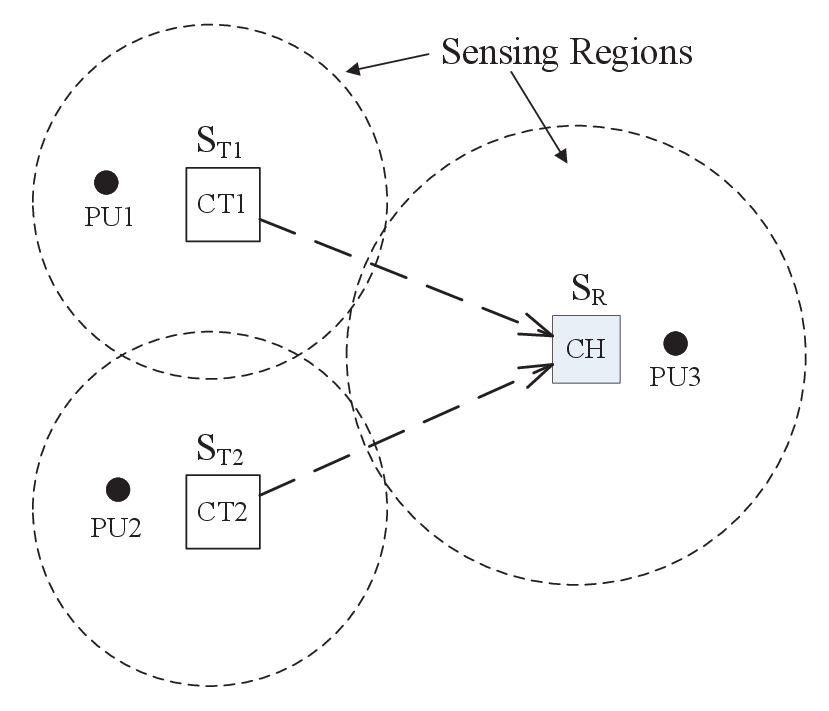}
\caption {Memoryless Cognitive MAC Channel} \label{Cog_MAC}
\end{figure}

\begin{figure}[t] \centering
\includegraphics[width=0.4\textwidth]{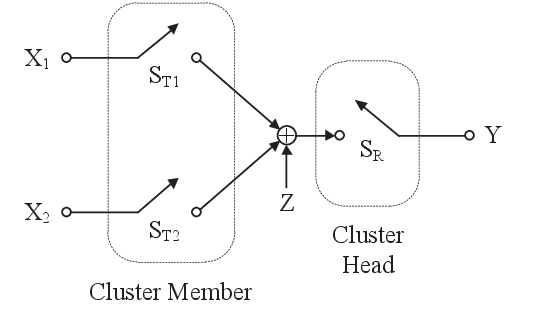}
\caption {Three Switch MAC Channel} \label{Three_Switch}
\end{figure}

The cognitive MAC channel with neighboring primary activities is
illustrated in Fig.~\ref{Cog_MAC}. We follow \cite{Two-switch} and model the memoryless cognitive MAC channel as a three-switch equivalent channel, as in Fig.~\ref{Three_Switch}. The states of switches at the two cognitive transmitters CT1, CT2 and the cognitive receiver CH are denoted respectively as $S_{T_1 }$, $S_{T_2 }$ and $S_R$, taking values of either $1$(on) or $0$(off). When the cognitive users detect (strong enough) interference from PU signals, the switch is turned off to avoid collisions. Otherwise, the switch is turned on for opportunistic communications. Based on this model, the input and output of the channel is related as:
\begin{equation}\label{cogMAC}
\begin{array}{l}
Y = \left( {{S_{{T_1}}}{X_1} + {S_{{T_2}}}{X_2} + Z} \right){S_R},\\
{S_{{T_1}}},{S_{{T_2}}},{S_R} \in \left\{ {0,1} \right\},
\end{array}
\end{equation}
where $X_1$ and $X_2$ are the transmitted symbols of CT1 and CT2
with average power constraint ${\rm{E}}\left[ {{{\left| {{X_i}}
\right|}^2}} S_{T_i} \right] \le {P_i}, i=1,2$, and $Z$ is the AWGN
noise with unit variance.

In this work, we assume \emph{perfect spectrum sensing} for ease of discussion. Then, the state of switch is actually controlled by the PU occupation, which is regarded as a type of \emph{side information} to the cognitive users. The side information is said to be causal if the transmitters or receiver only has knowledge of the past/current states of the corresponding switch, and non-causal if the future states are also known.

\section{Main Results}
\subsection{Achievable Rate and Capacity Regions of Cognitive MAC Channel}
We first explore the rate regions of the memoryless cognitive MAC
channel with independent transmitter side information, either causal or non-causal. Note that the causality of the receiver side
information does not matter in the study, as the receiver can decode after the transmission is finished. In the special case that the
transmitter side information is also known at the receiver, we can
derive the capacity region. These results are natural extension of
those in \cite{Two-switch} concerning the capacity of a single-link
two-switch channel.

1) Causal Side Information at the Transmitters. \\
\emph{Theorem 1:} For the three-switch channel with independent\footnote{Here, we consider the cases where
the side information of two transmitters are independent. Note this is reasonable when PU signal power is relatively low and the
cognitive transmitters keep a non-trivial distance with each other.} \emph{causal} side information $S_{T_1 }$ and $S_{T_2 }$
at the transmitters and side information $S_R$ at the receiver, coding can be performed directly on the input alphabets
(i.e., $ U_1 = X_1 $, $ U_2  = X_2 $ ) and an achievable rate
region is given by:
\begin{equation}
\begin{aligned}
&{\cal R}_{{S_{{T_1}}},{S_{{T_2}}},{S_R}}^{{\rm{causal}}} = \\
&\bigcup {\left\{ {\left( {{R_1},{R_2}} \right):\begin{array}{*{20}{c}}
{{R_1} \le I\left( {{X_1};Y,{S_R}\left| {{X_2}} \right.} \right)}\\
{{R_2} \le I\left( {{X_2};Y,{S_R}\left| {{X_1}} \right.} \right)}\\
{{R_1} + {R_2} \le I\left( {{X_1},{X_2};Y,{S_R}} \right)}
\end{array}} \right\}},
\end{aligned}
\end{equation}
for all $p\left( {\left. {{X_1},{X_2}} \right|{S_{{T_1}}},{S_{{T_2}}}} \right) = p\left( X_1 \right)p\left( X_2 \right)$, where $\bigcup$ denotes the convex hull of all rate pairs.

\emph{Remarks: } The above result indicates that in a white-space (on-off switching) multiple access channel, the optimal input distribution takes the form as if the channels were always on, like in general MAC. The achievable rate region is solely limited by the channel on-off behavior (reflected in the transition probabilities), and can be achieved by a code sequence with the optimal independent distribution after being masked at each ``off'' state.

2) Non-causal Side Information at the Transmitters. \\
\emph{Theorem 2:} For the three-switch channel with independent \emph{non-causal} side information $S_{T_1 }$ and $S_{T_2 }$ at the transmitters and side information $S_R$ at the receiver, coding can be performed directly on the input alphabets (i.e., $ U_1 = X_1 $, $U_2  = X_2 $ ) and an achievable rate region is given by:
{
\begin{equation}
\begin{aligned}
&{\cal R}_{{S_{{T_1}}},{S_{{T_2}}},{S_R}}^{{\rm{non - causal}}} =\\
& \bigcup
\left\{
    \left( {{R_1},{R_2}} \right):
        \begin{array}{*{20}{c}}
        {{R_1} \le I\left( {{X_1};Y,{S_R}\left| {{X_2}} \right.} \right) - I\left( {{X_1};{S_{{T_1}}}} \right)}\\
        {{R_2} \le I\left( {{X_2};Y,{S_R}\left| {{X_1}} \right.} \right) - I\left( {{X_2};{S_{{T_2}}}} \right)}\\
        {{R_1} + {R_2} \le \left(
            \begin{array}{c}
            I\left( {{X_1},{X_2};Y,{S_R}} \right)\\
            - I\left( {{X_1};{S_{{T_1}}}} \right) \\
            - I\left( {{X_2};{S_{{T_2}}}} \right)
            \end{array}
        \right)}
        \end{array}
\right\}.
\end{aligned}
\end{equation}}
for all $p\left( {\left. {{X_1},{X_2}} \right|{S_{{T_1}}},{S_{{T_2}}}} \right) = p\left( {\left. {{X_1}} \right|{S_{{T_1}}}} \right)p\left( {\left. {{X_2}} \right|{S_{{T_2}}}} \right)$.

\emph{Remarks: } In the non-causal case, the input distribution depends on the on-off side information, which indicates that the side information non-causally available at each transmitter now can be used to choose an optimal code sequence capable of adapting itself to the channel on-offs and facilitating the cognitive receiver's decoding with the side information which is possibly correlated to that of the transmitter. Intuitively, this will result in better achievable rates.

3) Strong Spectrum Sensing Capability at the Receiver. \\
\emph{Theorem 3:} For the three-switch channel, if the
transmitters' side information $S_{T_1 }$ and $S_{T_2
}$ are known to the receiver, i.e. $\left( {S_{T_1 } ,S_{T_2 }
} \right) = f\left( {S_R } \right)$, no matter whether the transmitters' side
information is \emph{causal or non-causal}, the channel capacity
regions are the same, as given by:
\begin{equation}
\begin{aligned}
&C_{\left( {{S_{{T_1}}},{S_{{T_2}}}} \right) = f\left( {{S_R}} \right)}^{{\rm{causal}},\;{\rm{non - causal}}}{\rm{ }} = \\
&\bigcup {\left\{ {\left( {{R_1},{R_2}} \right):\begin{array}{*{20}{c}}
{{R_1} \le I\left( {{X_1};Y,{S_R}\left| {{X_2}} \right.} \right)}\\
{{R_2} \le I\left( {{X_2};Y,{S_R}\left| {{X_1}} \right.} \right)}\\
{{R_1} + {R_2} \le I\left( {{X_1},{X_2};Y,{S_R}} \right)}
\end{array}} \right\}}.
\end{aligned}
\end{equation}
for all $p\left( {\left. {{X_1},{X_2}} \right|{S_{{T_1}}},{S_{{T_2}}}} \right) = p\left( {\left. {{X_1}} \right|{S_{{T_1}}}} \right)p\left( {\left. {{X_2}} \right|{S_{{T_2}}}} \right)$.

\emph{Remarks: } The above result indicates that, when the cognitive receiver has full knowledge about the transmitter on-off states, the optimal coding at each transmitter shall rely on the channel states, and the decoding at the receiver may make full use of those information. It is easy to see that the achievable rates are even better than those of the non-causal case as described in Theorem 2.

\subsection{Outer bounds and Inner bounds for Gaussian Switch Channel}
Since the above capacity regions are given implicitly in the form of mutual information, we further explore explicit outer and inner
bounds of the capacity region with the help of additional side information. We restrict to the Gaussian case to obtain the optimal results \cite{Two-switch}.

\emph{Definition 1:} To facilitate the analysis, we define six
events ($a$ to $f$) related to primary states $S_{T_1 }$, $S_{T_2 }$
and $S_R$, together with their probabilities of occurrence as
follows:
\begin{equation*}
\begin{array}{l}
a:\{ {S_R} = {S_{{T_1}}} = 1,{S_{{T_2}}} = 0\},~{p_a} \buildrel \Delta \over = p\left( {{S_R} = {S_{{T_1}}} = 1,{S_{{T_2}}} = 0} \right);\\
b:\{ {S_R} = {S_{{T_2}}} = 1,{S_{{T_1}}} = 0\},~{p_b} \buildrel \Delta \over = p\left( {{S_R} = {S_{{T_2}}} = 1,{S_{{T_1}}} = 0} \right);\\
c:\{ {S_R} = {S_{{T_1}}} = {S_{{T_2}}} = 1\} ,~{p_c} \buildrel \Delta \over = p\left( {{S_R} = {S_{{T_1}}} = {S_{{T_2}}} = 1} \right);\\
d:\{ {S_{{T_1}}} = 1,{S_{{T_2}}} = 0\} ,~{p_d} \buildrel \Delta \over = p\left( {{S_{{T_1}}} = 1,{S_{{T_2}}} = 0} \right);\\
e:\{ {S_{{T_1}}} = 0,{S_{{T_2}}} = 1\} ,~{p_e} \buildrel \Delta \over = p\left( {{S_{{T_1}}} = 0,{S_{{T_2}}} = 1} \right);\\
f:\{ {S_{{T_1}}} = {S_{{T_2}}} = 1\} ,~{p_f} \buildrel \Delta \over = p\left( {{S_{{T_1}}} = {S_{{T_2}}} = 1} \right).
\end{array}
\end{equation*}

Besides, we use $P_i^{j}$ to denote the power allocated to
user $i$ during event $j$.

\emph{Theorem 4:} For the three-switch MAC channel,
\emph{outer bound 1} of the capacity region can be obtained by assuming global side information at both the transmitters and the receiver:

{\begin{equation}\label{Theorem4}
\begin{aligned}
&C_{*,*,*}\left( {{P_1},{P_2}} \right) = \mathop  \bigcup \limits_{\scriptstyle
{{p_a}P_1^a + {p_c}P_1^c \le {P_1}} \hfill \atop
\scriptstyle {{p_b}P_2^b + {p_c}P_2^c \le {P_2}}
\hfill} \\
&\left\{ {\left( {{R_1},{R_2}} \right):
\begin{array}{*{20}{c}}
    {{R_1} \le {p_a}\log \left( {1 + P_1^a} \right) + {p_c}\log \left( {1 + P_1^c} \right)}\\
    {{R_2} \le {p_b}\log \left( {1 + P_2^b} \right) + {p_c}\log \left( {1 + P_2^c} \right)}\\
    {{R_1} + {R_2} \le \left(
    {\begin{array}{*{20}{c}}
        {p_a}\log \left( {1 + P_1^a} \right) \\
        + {p_b}\log \left( {1 + P_2^b} \right)\\
        + {p_c}\log \left( {1 + P_1^c + P_2^c} \right)
    \end{array}} \right)}
\end{array}} \right\}.
\end{aligned}
\end{equation}}

\emph{Theorem 5:} For the three-switch MAC channel, \emph{outer bound 2} of the capacity region can be obtained by assuming full side information only at the receiver:

{\begin{equation}\label{Theorem5}
\begin{aligned}
&C_{S_{T_1 } ,S_{T_2 } , * } \left( {P_1 ,P_2 }
\right) = \mathop  \bigcup \limits_{\scriptstyle P_1^d  = P_1^f  \le
\frac{{P_1 }}{{p_d  + p_f }} \hfill \atop
  \scriptstyle P_2^e  = P_2^f  \le \frac{{P_2 }}{{p_e  + p_f }} \hfill} \\
&\left\{
\left( {{R_1},{R_2}} \right):
    \begin{array}{*{20}c}
        R_1  \le p_a \log \left( {1 + P_1^d } \right) + p_c \log \left( {1 + P_1^f } \right)  \\
        R_2  \le p_b \log \left( {1 + P_2^e } \right) + p_c \log \left( {1 + P_2^f } \right)  \\
        R_1  + R_2  \le \left(
            \begin{array}{*{20}c}
                p_a \log \left( {1 + P_1^d } \right) \\
                + p_b \log \left( {1 + P_2^e } \right)  \\
                + p_c \log \left( {1 + P_1^f  + P_2^f } \right)
            \end{array} \right)\\
    \end{array} \right\}.
\end{aligned}
\end{equation}}

%\emph{Theorem 6.} \emph{outer bound 3}: For three-switch MAC
%channel with causal side information at the transmitters,
%\emph{outer bound 3} of capacity region can be obtained by assuming
%full side information only at both the transmitters. We find that
%\emph{outer bound 3} coincides with \emph{outer bound 1}, i.e.
%\begin{equation}
%C_{*,*,S_R }^{{\rm{causal}}} \left( {P_1 ,P_2 } \right) = C_{ * , *
%, * }^{{\rm{causal}}} \left( {P_1 ,P_2 } \right)
%\end{equation}

\emph{Theorem 6:} For the three-switch MAC channel
(with either causal or non-causal side information), an
\emph{inner bound} of the capacity region can be obtained as follows:
\begin{equation}
\begin{aligned}
&C_{{S_{{T_1}}},{S_{{T_2}}},{S_R}}^{{\rm{inner}}} =  \mathop \bigcup \limits_{\scriptstyle P_1^d  = P_1^f  \le
\frac{{P_1 }}{{p_d  + p_f }} \hfill \atop
  \scriptstyle P_2^e  = P_2^f  \le \frac{{P_2 }}{{p_e  + p_f }} \hfill} \\
&\left\{ {\left( {{R_1},{R_2}} \right):
    \begin{array}{*{20}{c}}
        R_1 \le {R_1^* - \Delta {R_1}}\\
        R_2  \le {R_2^* - \Delta {R_2}}\\
        R_1 + R_2 \le {R_1^* + R_2^* - \Delta \left( {{R_1} + {R_2}} \right)}
    \end{array}} \right\},
\end{aligned}
\end{equation}
where $\left(R_1^*, R_2^*\right)$ denotes the rate pair in \emph{outer bound 2}, and $0 \leq \Delta {R_1} \le {p_a}H\left( {\left. {{S_{{T_1}}}} \right|{S_R}} \right)$, $0 \leq \Delta {R_2} \le {p_b}H\left( {\left. {{S_{{T_2}}}} \right|{S_R}} \right)$, and $0 \leq \Delta \left( {{R_1} + {R_2}} \right) \le {p_c}H\left( {\left. {{S_{{T_1}}},{S_{{T_2}}}} \right|{S_R}} \right)$ denote the rate gaps.

\emph{Remarks: }When the PU states change very slowly, or are highly correlated among transmitters and receiver, it is shown that the outer bounds and inner bound coincide. In the case
with global side information, we can employ rate and power
allocation strategies to improve the system's overall performance.
Specifically, if we impose power constraints on both transmitters,
an optimal rate and power allocation scheme is obtained to maximize
the sum rate. We also analyze the effect of correlation in side information and PU occupation probability on the sum rate.

\section{Analysis}

\subsection{Preliminaries}
The memoryless cognitive MAC channel is modeled as a
three-switch channel, so that existing results on the memoryless MAC channel with
transmitter and receiver side information \cite{MAC-side} can be employed to derive
the cognitive rate regions, which are listed as the following lemmas:

\emph{Lemma 1.} \emph{Causal Case}: An achievable rate region of the discrete
memoryless MAC channel with receiver side information and
independent causal transmitter side
information is given by the convex closure of the rate pairs satisfying:
\begin{equation}\label{lemma1}
\bigcup\limits_{{p_{{\rm{causal}}}}} {\left\{ {\left( {{R_1},{R_2}} \right):\begin{array}{*{20}{c}}
{\begin{aligned}
{R_1} &\le I\left( {{U_1};Y,{S_R}\left| {{U_2}} \right.} \right) \\
&= I\left( {{U_1};Y\left| {{U_2},{S_R}} \right.} \right)\\
{R_2} &\le I\left( {{U_2};Y,{S_R}\left| {{U_1}} \right.} \right) \\
&= I\left( {{U_2};Y\left| {{U_1},{S_R}} \right.} \right)\\
{R_1} + {R_2} &\le I\left( {{U_1},{U_2};Y,{S_R}} \right) \\
&= I\left( {{U_1},{U_2};Y\left| {{S_R}} \right.} \right)
\end{aligned}}
\end{array}} \right\}},
\end{equation}
where the message is contained in the mutually independent auxiliary random
variables $U_1$ and $U_2$, and the causality is embodied in the following
conditional probability distribution:
$$ p_{{\rm{causal}}}  = \left\{ \begin{array}{l}
 p\left( {U_1 ,U_2 ,X_1 ,X_2 \left| {S_{T_1 } ,S_{T_2 } } \right.} \right) \\
  = p\left( {U_1 ,X_1 \left| {S_{T_1 } } \right.} \right)p\left( {U_2 ,X_2 \left| {S_{T_2 } } \right.} \right) \\
  = p\left( {U_1 } \right)p\left( {X_1 \left| {U_1 ,S_{T_1 } } \right.} \right)p\left( {U_2 } \right)p\left( {X_2 \left| {U_2 ,S_{T_2 } } \right.} \right) \\
 \end{array} \right\}.
$$
\emph{Lemma 2.} \emph{Non-causal Case}: An achievable rate region of the discrete
memoryless MAC channel with receiver side information and
independent non-causal transmitter side information is given by the convex closure of the rate pairs satisfying:
\begin{equation}\label{lemma2}
\bigcup\limits_{{p_{{\rm{non-causal}}}}} \left\{ {\left( {{R_1},{R_2}} \right):\begin{array}{*{20}{c}}
\begin{aligned}
{R_1} &\le \left( \begin{array}{l}
    I\left( {{U_1};Y,{S_R}\left| {{U_2}} \right.} \right) \\
    - I\left( {{U_1};{S_{{T_1}}}} \right)\\
    \end{array}\right)\\
{R_2} &\le \left( \begin{array}{l}
    I\left( {{U_2};Y,{S_R}\left| {{U_1}} \right.} \right) \\
    - I\left( {{U_2};{S_{{T_2}}}} \right)\\
     \end{array}\right)\\
{R_1} + {R_2} &\le \left( \begin{array}{l}
    I\left( {{U_1},{U_2};Y,{S_R}} \right) \\
    - I\left( {{U_1};{S_{{T_1}}}} \right) \\
    - I\left( {{U_2};{S_{{T_2}}}} \right)
     \end{array}\right)
\end{aligned}
\end{array}} \right\},
\end{equation}
where the message is contained in the mutually independent auxiliary random
variables $U_1$ and $U_2$, and the non-causality is embodied in the following
conditional probability distribution:
$$p_{{\rm{non - causal}}}  =
\left\{ \begin{array}{l}
 p\left( {U_1 ,U_2 ,X_1 ,X_2 \left| {S_{T_1 } ,S_{T_2 } } \right.} \right) \\
  = p\left( {U_1 ,X_1 \left| {S_{T_1 } } \right.} \right)p\left( {U_2 ,X_2 \left| {S_{T_2 } } \right.} \right) \\
  = \left(\begin{array}{l}
        p\left( {U_1 \left| {S_{T_1 } } \right.} \right)p\left( {X_1 \left| {U_1 ,S_{T_1 } } \right.} \right)\\
        \cdot~ p\left( {U_2 \left| {S_{T_2 } } \right.} \right)p\left( {X_2 \left| {U_2 ,S_{T_2 } } \right.} \right) \\
    \end{array} \right)
 \end{array} \right\}.
$$
\emph{Lemma 3.} If the transmitters' side information can be
expressed as a function of the receiver side information, i.e. $\left\{ {S_{T_1 } ,S_{T_2 } } \right\} = f\left( {S_R }
\right)$, the memoryless MAC channel capacity regions are the same for causal and
non-causal cases, i.e. $C^{{\rm{causal}}}  = C^{{\rm{non - causal}}}$, given by
\begin{equation}\label{lemma3}
\bigcup \left\{ {\left( {{R_1},{R_2}} \right):\begin{array}{*{20}{c}}
{{R_1} \le I\left( {{X_1};Y\left| {{S_R},{X_2}} \right.} \right)}\\
{{R_2} \le I\left( {{X_2};Y\left| {{S_R},{X_1}} \right.} \right)}\\
{{R_1} + {R_2} \le I\left( {{X_1},{X_2};Y\left| {{S_R}} \right.} \right)}
\end{array}} \right\},
\end{equation}
with the conditional probability distribution $p\left( {X_1 ,X_2 \left| {S_{T_1 } ,S_{T_2 } }
\right.} \right) = p\left( {X_1 \left| {S_{T_1 } } \right.}
\right)p\left( {X_2 \left| {S_{T_2 } } \right.} \right).$

\subsection{Achievable Rate and Capacity Regions with Causal and Non-causal Side Information}
In Shannon's seminal work on channel with side information \cite{SideInfom}, and later in Cover and Chiang's work on channel duality \cite{determ}, it is shown that it suffices for the optimal input of a channel with causal/non-causal side information to be a deterministic function of the side information and the auxiliary random variables. In \cite{MAC-side}, Jafar provides a unified view for both the causal and non-causal cases, and extends the results to the multiple access channel wherein independent side information at the two transmitters are assumed. Here in our study, by assuming independent on-off switching at the two transmitters, the above sufficient condition for the optimal inputs to be deterministic functions of the side information and auxiliary variables is satisfied, which lays a basis for the following proofs.

\begin{proof}[1) Proof for Theorem 1]
As stated in \cite{determ} and \cite{MAC-side}, it suffices for the optimal inputs $X_1$, $X_2$ to be deterministic functions of the causal side information $S_{T_1 }$, $S_{T_2 }$ and the auxiliary random variables $U_1$, $U_2$, i.e., $X_1  = f_1 \left( {U_1 ,S_{T_1 } } \right)$ and $X_2  = f_2 \left( {U_2 ,S_{T_2 } } \right)$. Since $ S_{T_1 } ,S_{T_2 }  \in \left\{ {0,1} \right\}$, we define
$$
\begin{aligned}
{X_1} = \left\{ {\begin{array}{*{20}{c}}
{{g_1}\left( {{U_1}} \right),\quad {\rm{for}}\;{S_{{T_1}}} = 1},\\
{{h_1}\left( {{U_1}} \right),\quad {\rm{for}}\;{S_{{T_1}}} = 0},\\
\end{array}} \right. \\
{X_2} = \left\{ {\begin{array}{*{20}{c}}
{{g_2}\left( {{U_2}} \right),\quad {\rm{for}}\;{S_{{T_2}}} = 1},\\
{{h_2}\left( {{U_2}} \right),\quad {\rm{for}}\;{S_{{T_2}}} = 0}.\\
\end{array}} \right.
\end{aligned}
$$

Note that in the switch channel, when $S_{T_1 }$ or $S_{T_2 }$ is zero, $Y$ is not influenced by
$X_1$ or $X_2$ at all. Therefore, it is reasonable to assume
$$
\begin{array}{l}
 X_1  = g_1 \left( {U_1 } \right),\quad {\rm{for}} \ S_{T_1 }  = 0,1, \\
 X_2  = g_2 \left( {U_2 } \right),\quad {\rm{for}} \ S_{T_2 }  = 0,1. \\
\end{array}
$$
That means it suffices for the optimal $X_i \left( i=1, 2\right)$ to be a deterministic function of $U_i$, in other words, decoding $U_i $ is sufficient for decoding $X_i$. On the other hand, since the information rate transmitted over a channel with causal side information is bounded by the maximal mutual information between the channel output and the auxiliary variable \cite{SideInfom, MAC-side}, it is necessary for the channel input $X_i$ to carry all information $U_i$ may have so as to achieve the maximal rate. Consequently, decoding $X_i$ should also be sufficient for recovering information in $U_i$. The above facts reveal that $U_i$ and $X_i$ are in fact equivalent in the sense of carrying information. Therefore, without loss of generality, the rate region of cognitive MAC channel can be formulated by replacing $U_1$ and $U_2$ with $X_1$ and $X_2$ respectively in \emph{Lemma 1}.

Furthermore, $U_1 \left( {U_2 } \right)$ is independent of $S_{T_1 } \left({S_{T_2 } } \right)$ according to Shannon's coding theorem for channel with causal side information \cite{SideInfom} (see also \emph{Lemma 1}). As a result, it suffices for the input distribution to take the form of
$$p\left( {\left.
{{X_1},{X_2}} \right|{S_{{T_1}}},{S_{{T_2}}}} \right) =
p(X_1)p(X_2).$$

\end{proof}

\hspace{1sp}
\begin{proof}[2) Proof for Theorem 2] We first derive the rate of
$R_1$ based on \emph{Lemma 2}.
\begin{align*}
{R_1} &\le I\left( {{U_1};Y,{S_R}\left| {{U_2}} \right.} \right) - I\left( {{U_1};{S_{{T_1}}}} \right) \\
& \mathop = \limits^{(a)} I\left( {U_1 ,X_1 ;Y,S_R \left| {U_2 ,X_2 } \right.} \right) - I\left( {U_1 ,X_1 ;S_{T_1 } } \right) \\
 & = \left( \begin{array}{l}
 I\left( {X_1 ;Y,S_R \left| {U_2 ,X_2 } \right.} \right) + I\left( {U_1 ;Y,S_R \left| {X_1 ,U_2 ,X_2 } \right.} \right) \\
  - \left( {I\left( {X_1 ;S_{T_1 } } \right) + I\left( {U_1 ;S_{T_1 } \left| {X_1 } \right.} \right)} \right) \\
 \end{array} \right) \\
& \mathop  = \limits^{(b)} \left( \begin{array}{l}
 I\left( {X_1 ;Y,S_R \left| {X_2 } \right.} \right) - I\left( {X_1 ;S_{T_1 } } \right) \\
  + I\left( {U_1 ;Y,S_R \left| {X_1 } \right.} \right) - I\left( {U_1 ;S_{T_1 } \left| {X_1 } \right.} \right) \\
 \end{array} \right) \\
& \mathop  \le \limits^{(c)} I\left( {X_1 ;Y,S_R \left| {X_2 } \right.} \right) - I\left( {X_1 ;S_{T_1 } } \right),
\end{align*}
{\flushleft{where}} \\
\indent $(a)$ is due to $ X_i  = g_i \left( {U_i } \right), \
{\rm{for}} \ S_{T_i }  = 0,1, i=1, 2$.
%Note that, here the expression only means that in the switch channel $X_i$ can be solely determined by the variable $U_i$ which is conditioned on $S_{T_i}$ (see \emph{Lemma 2}), i.e., $X_i$ is still conditioned on $S_{T_i}$.
Note that from \emph{Lemma 2}, $U_i$ is conditioned on $S_{T_i}$, that means $X_i$ is still conditioned on $S_{T_i}$ but can be solely determined by the variable $U_i$ in the switch channel;\\
\indent $(b)$ holds as $S_{T_1}$ and $S_{T_2}$ are independent, thus
$U_2,X_2$ are independent of $U_1,X_1$; \\
\indent $(c)$ is obtained as $ \left( {U_1 \left| {X_1  = x_1 } \right.} \right)
\to \left( {S_{T_1 } \left| {X_1  = x_1 } \right.} \right) \to
\left( {Y,S_R \left| {X_1  = x_1 } \right.} \right)$ forms a Markov
Chain, or equivalently, $ I\left( {U_1 ;Y,S_R \left| {X_1 } \right.}
\right) - I\left( {U_1 ;S_{T_1 } \left| {X_1 } \right.} \right) \le
0$, and the equality is achievable when simply setting $U_1=X_1$.

By symmetry we can similarly obtain $ R_2  \le \mathop {\max
}\limits_{p\left( {X_2 \left| {S_{T_2 } } \right.} \right)} I\left(
{X_2 ;Y,S_R \left| {X_1 } \right.} \right) - I\left( {X_2 ;S_{T_2 }
} \right) $.\vspace{3mm}

For the sum rate of $R_1+R_2$, from \cite{MAC-side}, we have
\begin{align}
 R_1  + R_2  &\le \left( \begin{array} {l}
    I\left( {U_1 ,U_2 ;Y,S_R } \right)\\
    - I\left( {U_1 ;S_{T_1 } } \right) - I\left( {U_2 ;S_{T_2 } } \right)
    \end{array} \right) \label{1st-line-sumrate} \\ \nonumber
 &\mathop  = \limits^{\left( a \right)} \left( \begin{array} {l}
    I\left( {U_1 ,X_1 ,U_2 ,X_2 ;Y,S_R } \right) \\
    - I\left( {U_1 ,X_1 ;S_{T_1 } } \right) - I\left( {U_2 ,X_2 ;S_{T_2 } } \right)
    \end{array} \right) \\ \nonumber
 & = \left( \begin{array}{l}
    I\left( {X_1 ,X_2 ;Y,S_R } \right) + I\left( {U_1 ,U_2 ;Y,S_R \left| {X_1 ,X_2 } \right.} \right) \\
    - \left( {I\left( {X_1 ;S_{T_1 } } \right) + I\left( {U_1 ;S_{T_1 } \left| {X_1 } \right.} \right)} \right) \\
    - \left( {I\left( {X_2 ;S_{T_2 } } \right) + I\left( {U_2 ;S_{T_2 } \left| {X_2 } \right.} \right)} \right) \\
    \end{array} \right)  \\ \nonumber
 &\mathop  = \limits^{\left( b \right)} \left( \begin{array}{l}
    I\left( {X_1 ,X_2 ;Y,S_R } \right) - I\left( {X_1 ;S_{T_1 } } \right) \\
    - I\left( {X_2 ;S_{T_2 } } \right) + I\left( {U_1 ,U_2 ;Y,S_R \left| {X_1 ,X_2 } \right.} \right) \\
    - I\left( {U_1 ;S_{T_1 } \left| {X_1 ,X_2 } \right.} \right) - I\left( {U_2 ;S_{T_2 } \left| {X_1 ,X_2 } \right.} \right) \\
    \end{array} \right)  \\ \nonumber
 &\mathop  = \limits^{\left( c \right)} \left( \begin{array}{l}
    I\left( {X_1 ,X_2 ;Y,S_R } \right) - I\left( {X_1 ;S_{T_1 } } \right) \\
    - I\left( {X_2 ;S_{T_2 } } \right) + I\left( {U_1 ,U_2 ;Y,S_R \left| {X_1 ,X_2 } \right.} \right) \\
    - I\left( {U_1 ,U_2 ;S_{T_1 } ,S_{T_2 } \left| {X_1 ,X_2 } \right.} \right) \\
    \end{array} \right)  \\
 &\mathop  \le \limits^{\left( d \right)} {I\left( {X_1 ,X_2 ;Y,S_R } \right) - I\left( {X_1 ;S_{T_1 } } \right) - I\left( {X_2 ;S_{T_2 } } \right)} \label{last-line-sumrate},
\end{align}
{\flushleft{where}}\\
\indent $(a)$ is due to $ X_i  = g_i \left( {U_i } \right), \ {\rm{for}} \ S_{T_1 }  = 0,1, \ i = 1,2$;\\
\indent $(b)$ and $(c)$ hold as $S_{T_1}$ and $S_{T_2}$ are independent, thus $U_1,X_1$ are independent of $U_2,X_2$;\\
%$c$ is because $ I\left( {A,B;C,D} \right) \le I\left( {A;C} \right)
%+ I\left( {B;D} \right) $.
%$c$ is because $S_{T_1}$ and $S_{T_2}$, $U_1$ and $U_2$ are
%independent.
\indent $(d)$ is obtained as $ ( {U_1 | X_1  = x_1 ,X_2  = x_2
} ) \to ( S_{T_1 } | {X_1  = x_1 ,X_2  = x_2
}  ) \to ( Y,S_R | X_1  = x_1 ,X_2  = x_2
) $ forms a Markov chain, so does $ ( {U_2 | {X_1  = x_1 ,X_2 =
x_2 } } ) \to ( S_{T_2 } | {X_1  = x_1 ,X_2
= x_2 } ) \to ( Y,S_R | {X_1  = x_1 ,X_2 =
x_2 }  )$. Hence, $ ( U_1 ,U_2 | X_1  = x_1 ,X_2  = x_2
)\to (S_{{T_1}},S_{{T_2}}|{X_1} = {x_1},{X_2} = {x_2}) \to ( Y,S_R | {X_1  = x_1 ,X_2  =
x_2 }  )$ also forms a Markov Chain, or equivalently, $I( {U_1 ,U_2 ;Y,S_R | {X_1 ,X_2 } } )\\
-
I( U_1 ,U_2 ;S_{T_1 } ,S_{T_2 } | {X_1 ,X_2 }
) \le 0 $. Finally, the equality in \eqref{last-line-sumrate} is achievable by setting $U_1=X_1$ and $U_2=X_2$ in \eqref{1st-line-sumrate}.
\end{proof}
\begin{proof}[3) Proof for Theorem 3]
%For the memoryless cognitive MAC channel, if the transmitters'
%neighboring primary activities $S_{T_1 }$ and $S_{T_2 }$ are also
%sensed by the receiver, it is modeled as the three-switch MAC
%channel with \emph{strong sensing ability at the CH}. This is
%equivalent to the case of \emph{Lemma 3}, where the transmitters'
%side information is a function of the receiver side information,
%i.e. $\left( {S_{T_1 } ,S_{T_2 } } \right) = f\left( {S_R } \right)$
% \cite{MAC-side}, which is also known to the receiver, .
Since the three-switch MAC channel is a special case of the
memoryless channel with side information, only that the side
information here is binary, we can directly employ the result of
\emph{Lemma 3} to complete the proof.
\end{proof}
\emph{Remarks:} The assumption in \emph{Theorem 3}
requires that CH (or BS) knows the primary activities not only at
its own side, but also near the transmitters. In practice, it may not be easy for the CH (or BS) to obtain this knowledge directly. But in some cases, with the aid of channel feedback or out-of-band signalling, it can be achieved indirectly.

\subsection{Outer and Inner Bounds}
Since the above capacity regions are given implicitly in mutual information, we further explore explicit outer and inner
bounds of the capacity region with the help of additional side information for the typical Gaussian channel \cite{Two-switch}.

We consider two kinds of additional side information at the cognitive transmitters and/or receiver, and derive the corresponding outer bounds.
In \emph{Case 1}, both the transmitters and the receiver have full
knowledge of all side information (i.e., $S_{T_1}$, $S_{T_2}$, and $S_R$). In \emph{Case 2}, only the receiver has full side information. For the case when only the transmitters (both) have full side information, either of  them transmits only when its own switch is on and $S_R=1$. We assume that the receiver can infer the states of transmitters through signal detection \footnote{With the assumption that the receiver knows the codebooks of both transmitters, the receiver can distinguish when there is one or two active transmitters.}, thus this case coincides with \emph{Case 1}.
Note that when the receiver also has transmitter side information, there is no differences between the causal and non-causal case \cite{Two-switch}. Also, arbitrary correlation among side information may be considered in these genie-aided scenarios.

\subsubsection{Outer bound for Case 1-Global Side Information}
With global side information, the transmissions occur when $S_R = 1$ and
$S_{T_1} + S_{T_2} \ge 1$, which can be further categorized into three subcases.

When $ S_R  = S_{T_1 }  = 1,S_{T_2 }  = 0$, the MAC channel degrades
into a point-to-point channel, in which Gaussian input is optimal. Assuming CT1's transmission power to
be $ P_1^a$, the achievable rate region corresponding to event $a$ is {${C_{ a, *, *, * } \left( {P_1^a ,0}
\right) = \bigcup \left\{ {\begin{array}{*{20}c}
   {R_1^a  \le \log \left( {1 + P_1^a } \right)}  \\
   {R_2^a  = 0}  \\
\end{array}} \right\}.}
$}
Similarly, when $ S_R  = S_{T_2 }  = 1,S_{T_1 }  = 0$, assuming
CT2's transmission power to be $ P_2^b$, the achievable rate region corresponding to event
$b$ is
{$C_{ b, * , * , * } \left( {0,P_2^b }
\right) =  \bigcup  \left\{ {\begin{array}{*{20}c}
   {R_1^b  = 0}  \\
   {R_2^b  \le \log \left( {1 + P_2^b } \right)}  \\
\end{array}} \right\}.$}
When $ S_R  = S_{T_1 }  = S_{T_2 }  = 1 $, the channel is the traditional
MAC channel. Assuming the transmission power of CT1 and CT2 to be $
P_1^c$ and $ P_2^c$, respectively, the achievable rate region corresponding to event $c$ is:

{$$C _{ c, * , * , * } \left( {P_1^c
,P_2^c } \right) =  \bigcup  \left\{ {\begin{array}{*{20}c}
   {R_1^c  \le \log \left( {1 + P_1^c } \right)}  \\
   {R_2^c  \le \log \left( {1 + P_2^c } \right)}  \\
   {R_1^c  + R_2^c  \le \log \left( {1 + P_1^c  + P_2^c } \right)}  \\
\end{array}} \right\}.$$}
Taking into account the power constraints: $
 p_a P_1^a  + p_c P_1^c  \le P_1 $, and $
 p_b P_2^b  + p_c P_2^c  \le P_2 $,
\emph{outer bounds 1} can be derived as in \eqref{Theorem4}.

\emph{Optimal Rate/Power Allocation:}
When both the transmitters and receiver have global state
information, we can further explore the optimal rate and power
allocation.

Without loss of generality, we may take the optimal sum rate as our objective:
\begin{equation}\label{optimization}
\begin{array}{l}
 {\rm{maximize:}}\;R_1  + R_2  = \left( \begin{array} {l}
    p_a \log \left( {1 + P_1^a } \right) + p_b \log \left( {1 + P_2^b } \right) \\
    + p_c \log \left( {1 + P_1^c  + P_2^c } \right)
    \end{array} \right), \\
 {\rm{subject~to:}}\;\left\{ \begin{array}{l}
 p_a P_1^a  + p_c P_1^c  \le P_1 , \\
 p_b P_2^b  + p_c P_2^c  \le P_2 , \\
 P_1^a ,P_2^b ,P_1^c ,P_2^c  \ge 0 .\\
 \end{array} \right. \\
 \end{array}
\end{equation}
This is a convex optimization problem, and can
be solved through KKT conditions.
$$ \left\{ \begin{array}{l}
 \frac{{\partial L\left( {P_1^a ,P_2^b ,P_1^c ,P_2^c } \right)}}{{\partial P_1^a }} = 0 ,\\
 \frac{{\partial L\left( {P_1^a ,P_2^b ,P_1^c ,P_2^c } \right)}}{{\partial P_2^b }} = 0 .\\
 \end{array} \right.
$$

Substituting the constraints $ P_1^c  = \frac{{P_1  - p_a P_1^a
}}{{p_c }},P_2^c  = \frac{{P_2  - p_b P_2^b }}{{p_c }} $, the
solution is obtained as
\begin{equation}\label{optimal-solution}
\begin{array}{l}
 P_1^a  = P_2^b  = \frac{{P_1  + P_2 }}{{p_a  + p_b  + p_c }} ,\\
 P_1^c  = \frac{{\left( {p_b  + p_c } \right)P_1  - p_a P_2 }}{{\left( {p_a  + p_b  + p_c } \right)p_c }} ,\\
 P_2^c  = \frac{{\left( {p_a  + p_c } \right)P_2  - p_b P_1 }}{{\left( {p_a  + p_b  + p_c } \right)p_c }} .\\
 \end{array}
\end{equation}

The corresponding optimal sum rate is given as follows.

\emph{Corollary 1:} The maximal sum rate is $ \left(R_1  + R_2\right)_{\rm{max}}  =
\left( {p_a + p_b  + p_c } \right)\log \left( {1 + \frac{{P_1  + P_2
}}{{p_a  + p_b  + p_c }}} \right) $.

\subsubsection{Outer bound for Case 2-Full Side Information at Receiver}
In \emph{Case 2}, transmissions occur only when $ S_{T_1 }  + S_{T_2 }  \ge 1 $. This can be further categorized into three subcases,
which correspond to the events $d$, $e$ and $f$ in \emph{Definition 1}.

When $ S_{T_1 }  = 1,S_{T_2 }  = 0$ or $ S_{T_1 }  = 0,S_{T_2 }  =
1$, the MAC channel degrades into a point-to-point channel. We assume
CT1 and CT2's transmission power to be $\left( {P_1^d ,0} \right)$
and $\left( {0, P_1^e} \right)$, respectively. When $S_R=1$, the
achievable rate regions for event $d$ and $e$ are {$ C_{d, S_{T_1 } ,S_{T_2 } , * } \left(
{P_1^d ,0} \right) = \bigcup \left\{ {\begin{array}{*{20}c}
   {R_1^d  \le \log \left( {1 + P_1^d } \right)}  \\
   {R_2^a  = 0}  \\
\end{array}} \right\}$} and {
$ C_{e, S_{T_1 } ,S_{T_2 } , * }
\left( {0,P_2^e } \right) = \bigcup \left\{ {\begin{array}{*{20}c}
   {R_1^e  = 0}  \\
   {R_2^e  \le \log \left( {1 + P_2^e } \right)}  \\
\end{array}} \right\}
$}.

When $ S_{T_1 }  = S_{T_2 }  = 1 $, the channel becomes the basic
MAC channel. We assume the transmission power of CT1 and CT2 to be $P_1^f$ and $P_2^f$, respectively.  When $S_R=1$, the achievable rate region for event $f$ is:

$$ \begin{aligned}
&C_{f, S_{T_1 } ,S_{T_2 } , * } \left({P_1^f ,P_2^f } \right) = \\
&\bigcup \left\{ {\begin{array}{*{20}c}
   {R_1^f  \le \log \left( {1 + P_1^f } \right)}  \\
   {R_2^f  \le \log \left( {1 + P_2^f } \right)}  \\
   {R_1^f  + R_2^f  \le \log \left( {1 + P_1^f  + P_2^f } \right)}  \\
\end{array}} \right\},
\end{aligned}$$
where the power constraints are $p_d P_1^d  + p_f P_1^f  \le P_1$ and $
 p_e P_2^e  + p_f P_2^f  \le P_2$.

Since the transmitters do not have full side information, they can
not discriminate the three subcases and have to use the same
transmission power, i.e. $ P_1^d  = P_1^f$, $ P_2^e  = P_2^f$. The transmission power is bounded as $ P_1^d  = P_1^f  \le
\frac{{P_1 }}{{p_d  + p_f }}$ and $P_2^e  = P_2^f  \le \frac{{P_2
}}{{p_e  + p_f }}$.

Although transmissions occur with the probabilities of $p_d$, $p_e$
and $p_f$, effective transmissions occur only when $S_R = 1$, with
the probabilities of $p_a$, $p_b$ and $p_c$. Only effective
transmissions contribute to system capacity, therefore
\emph{outer bound 2} is derived as in \eqref{Theorem5}.

%\subsubsection{Outer bounds 3 - Full Side Information at Transmitters}
%In \emph{Case 3}, although the receiver (or CH) does not know the
%transmitters' side information, it can infer their states through
%detecting their transmitted signals. This is equivalent with
%\emph{Case 1}. Effective transmissions occur only when $ S_{T_1 }  +
%S_{T_2 } \ge 1 $ and $ S_R  = 1 $, thus \emph{outer bounds 3} is the
%same as \emph{outer bounds 1}, i.e. $ C_{*,*,S_R }^{{\rm{causal}}}
%\left( {P_1 ,P_2 } \right) = C_{
%* ,
%* , * }^{{\rm{causal}}} \left( {P_1 ,P_2 } \right) $, which proves
%\emph{Theorem 6}.

\subsubsection{Inner bound}
The inner bound can not be calculated directly. However, we can
obtain it with the help of genie information \cite{Two-switch}.
Suppose a genie provides some additional information about the transmitter states to the receiver
every channel use through a genie variable \textit{G}. The basic idea of
genie-aided lower bound is that ``the improvement in capacity
induced by the genie information \textit{G} cannot exceed the entropy rate of the genie information itself''.

When $ S_{T_1 }  = 1,S_{T_2 }  = 0$ or $ S_{T_1 } = 0,S_{T_2 } = 1$,
the MAC channel degrades to a point-to-point channel. Employing the genie-aided bound, the inner bounds for the two subcases are $
R_1 \ge R_1^* - H\left( {{G_1}\left| {{S_R}} \right.} \right)$ and $R_2 \ge R_2^* - H\left( {{G_2}\left| {{S_R}} \right.} \right)$.

When $ S_{T_1 }  = 1,S_{T_2 }  = 1$, it is now a MAC channel. Since for the MAC, the maximal sum rate improvement due to the availability of genie information is bounded by the amount of genie information itself \cite[Theorem 6]{MAC-side}, i.e., $ C^\Sigma_{S_{T_1 } ,S_{T_2 } , \left( S_R , G \right)}  - C^\Sigma_{S_{T_1 } ,S_{T_2 } ,S_R } \le H\left( {G\left| {S_R } \right.} \right)$, we thus get the lower bound on sum rate $R_1 + R_2 \ge R_1^* + R_2^* - H\left( {{G_1},{G_2}\left| {{S_R}} \right.} \right)$, where $\left(R_1^*, R_2^*\right)$ denotes the maximal rate pair in \emph{outer bound 2}.

To sum up, $\Delta {R_1} \le {p_a}H\left( {\left. {{G_1}} \right|{S_R}} \right) \le {p_a}H\left( {\left. {{S_{{T_1}}}} \right|{S_R}} \right)$, $\Delta {R_2} \le {p_b}H\left( {\left. {{G_2}} \right|{S_R}} \right) \le {p_b}H\left( {\left. {{S_{{T_2}}}} \right|{S_R}} \right)$, and $\Delta \left( {{R_1} + {R_2}} \right) \le {p_c}H\left( {\left. {{G_1},{G_2}} \right|{S_R}} \right) \le {p_c}H\left( {\left. {{S_{{T_1}}},{S_{{T_2}}}} \right|{S_R}} \right)$. Thus, the overall inner bound is obtained as in \emph{Theorem 6}.

\emph{Remarks:}
According to \emph{Theorem 6}, in event $d$, $e$ and $f$, the
inner bounds are $H\left( {G_1 \left| {S_R } \right.} \right)$,
$ H\left( {G_2 \left| {S_R } \right.} \right)$ and $ H\left(
{G_1  ,G_2 \left| {S_R } \right.} \right)$ bits per channel
use lower than the outer bounds with full side information at the receiver, respectively.

The transmitters only need to send one bit notification to the receiver when there is a change of PU states. We assume that, on average, the PU activities at the transmitters change every $N$ time slots. Then the genie
information rate is at most $1/N$ bit per time slot for each transceiver pair. Formally, the gap between \emph{outer bound 2} and \emph{inner bound} is restricted as $H\left( {\left. {{G_1}} \right|{S_R}} \right) = H\left( {\left. {{G_2}} \right|{S_R}} \right) \le H\left( {{G_1}} \right) = H\left( {{G_2}} \right) = \frac{1}{N}$ and $H\left( {\left. {{G_1},{G_2}} \right|{S_R}} \right) \le H\left( {{G_1},{G_2}} \right) = \frac{2}{N}$.

Moreover, when the side information between the transmitters and receiver are highly correlated, $ H\left( {S_{T_1 }
\left| {S_R } \right.} \right)$, $ H\left( {S_{T_2 } \left| {S_R } \right.} \right)$
and $ H\left( {S_{T_1 } ,S_{T_2 } \left| {S_R } \right.} \right)$ also converge to
zero, and so do $H\left( {\left. {{G_1}} \right|{S_R}} \right)$, $H\left( {\left. {{G_2}} \right|{S_R}} \right)$ and $H\left( {\left. {{G_1},{G_2}} \right|{S_R}} \right)$. Therefore, we conclude that the outer bounds and inner bound
coincide when the states of PU signals change very slowly or the side information between the transmitters and
receiver are highly correlated.

\subsection{Effect of correlation and PU occupation rate}
We take the optimized sum rate in \emph{Corollary 1} as an example, and analyze how it is influenced by two system parameters within a specific probabilistic model. We assume the PU occupation rates at both transmitters and the receiver to be the same, $ \mu  = 1 - E\left[ {S_{T_1 } } \right] =1- E\left[ {S_{T_2 } } \right] = 1-E\left[ {S_R } \right]$; and the mutual correlation coefficient between the three states also to be the same, as $\rho$. It is relatively straightforward to extend the study to the more general scenario regarding channel correlation and PU activity. According to the definition of correlation coefficient, the joint probability distribution is denoted in Table I, where $p_0 = \mu \left[ {{\mu ^2} + \rho \left( {1 - {\mu ^2}} \right)} \right]$,
$p_1 = \left( {1 - {\rho ^2}} \right)\left( {1 - \mu } \right){\mu ^2}$,
$p_2 = (1 - \rho )\mu \left[ {{{\left( {1 - \mu } \right)}^2} + \rho (\mu  - {\mu ^2})} \right]$, and $p_3 = \left[ {{{\left( {1 - \mu } \right)}^2} + \rho (\mu  - {\mu ^2})} \right](1 - \mu  + \mu \rho )$.
The probabilities for the six events in \emph{Definition 1} can be expressed as functions of $\mu$ and $\rho$:
\begin{equation}\label{probabilities}
\left\{ \begin{array}{l}
{p_a} = {p_b} = p_2, \quad \quad \quad {p_c} = p_3,\\
{p_d} = {p_e} = p_1 + p_2, \quad {p_f} = p_2 + p_3.
\end{array}  \right.\\
\end{equation}

\begin{table}[h]
\renewcommand{\arraystretch}{1.1}
\caption{Joint Probability Distribution under Non-fading and Fading Models} \label{table_example}
\centering
\begin{tabular}{|c|c|c|}
\hline
\multicolumn{3}{|c|}{$S_R=0$} \\
\hline
\backslashbox{${S_{{T_1}}}$}{${S_{{T_2}}}$} & $0$ & $1$ \\
\hline
$0$ & $p_0$ & $p_1$ \\
\hline
$1$ & $p_1$ & $p_2$ \\
\hline \hline
\multicolumn{3}{|c|}{$S_R=1$} \\
\hline
\backslashbox{${S_{{T_1}}}$}{${S_{{T_2}}}$} & $0$ & $1$ \\
\hline
$0$ & $p_1$ & $p_2$ \\
\hline
$1$ & $p_2$ & $p_3$ \\
\hline
\end{tabular}
\end{table}

Combining \eqref{optimal-solution} and \eqref{probabilities}, the
sum rate can be re-written in the forms of $\rho$ and $\mu$:
\begin{equation}\label{sumrate-mu-rho}
{\left( {{R_1} + {R_2}} \right)_{\max }} = p\left( {\mu ,\rho }
\right)\log \left( {1 + \frac{{{P_1} + {P_2}}}{{p\left( {\mu ,\rho }
\right)}}} \right),
\end{equation}
where $p\left( {\mu ,\rho } \right) = \left( {1 + \mu  - \mu \rho } \right)\left[ {{{\left( {1 - \mu } \right)}^2} + \rho (\mu  - {\mu ^2})} \right]$.

We find that \eqref{sumrate-mu-rho} is monotonically increasing with $p\left( {\mu ,\rho } \right)$. Moreover, since $0<\mu,\rho<1$,
$$\begin{array}{l}
\frac{{\partial p\left( {\mu ,\rho } \right)}}{{\partial \mu }} = \mu {\left( {1 - \rho } \right)^2}\left( {3\mu  - 2} \right) - 1 < 0,\\
\frac{{\partial p\left( {\mu ,\rho } \right)}}{{\partial \rho }} = 2{\mu ^2}\left( {1 - \rho } \right)\left( {1 - \mu } \right) > 0.
\end{array}$$

Thus \eqref{sumrate-mu-rho} is a monotonically decreasing function
of $\mu$ and monotonically increasing function of $\rho$. The
insights here are that the sum rate increases when the PU is less
active and when the correlation among the side information is
stronger.

\subsection{Extension to Interference and Fading model}
We first consider a natural extension of \eqref{cogMAC}, in which the secondary receiver keeps on all the time. It is reasonable as the receiving process will not cause interference to the PU system. Our intention is to examine how much gain we can obtain by allowing the secondary receiver to remain receiving even in the presence of PU interference. Thus the received signal falls in one of the following two categories:
\begin{equation}\label{InterferenceModel}
Y = \left\{ {\begin{array}{*{20}{c}}
{{S_{{T_1}}}{X_1} + {S_{{T_2}}}{X_2} + Z,\quad {S_R} = 1},\\
{{S_{{T_1}}}{X_1} + {S_{{T_2}}}{X_2} + I,\quad {S_R} = 0},
\end{array}} \right.
\end{equation}
where $I$ denotes the PU interference plus noise, which is modeled as a Gaussian variable with variance $P_I$. ${{S_{{T_1}}}}$, ${{S_{{T_2}}}}$ and ${{S_{{R}}}}$ are defined the same as in the three-switch model \eqref{cogMAC}.

Following the same line in \emph{Corollary 1} (Section IV.C), the maximal sum rate for
cognitive MAC channel with interference is obtained after solving the convex optimization problem:
\begin{equation}\label{max-sumrate-fading}
\begin{array}{l}
\begin{aligned}
&{\left( {{R_1} + {R_2}} \right)_{\max }} = \\
&\mathop { \max }\limits_{{{P'}_1} + {{P'}_2} + {{P''}_1} + {{P''}_2} \le {P_1} + {P_2}} \left[
    \begin{array} {l} {p_{{ni}}}\log \left( {1 + \frac{{{{P'}_1} + {{P'}_2}}}{{{p_{{ni}}}}}} \right) \\
    + {p_{{i}}}\log \left( {1 + \frac{{{{P''}_1} + {{P''}_2}}}{{{p_{{i}}}{P_I}}}} \right)\\
    \end{array} \right],
\end{aligned}
\end{array}
\end{equation}
where ${p_{{ni}}} \buildrel \Delta \over = {p_a} +
{p_b} + {p_c}$ and ${p_{{i}}} \buildrel \Delta \over =
{p_d} + {p_e} + {p_f} - \left( {{p_a} + {p_b} + {p_c}} \right)$ are the probabilities with and without PU interference at the receiver, respectively.
${P'_1} + {P'_2}$ is the power spent when ${S_R} = 1$, and ${P''_1}
+ {P''_2}$ is the power spent when ${S_R} = 0$, both of which are
under the average power constraints. It can be shown that the
above maximum is achieved when
\begin{equation}\label{max-solution-fading}
\begin{array}{l}
{{P''}_1} + {{P''}_2} = {\left( {\frac{{{p_{{i}}}\left( {{P_1} + {P_2}} \right) - {p_{{ni}}}{p_{{i}}}\left( {{P_I} - 1} \right)}}{{{p_{{ni}}} + {p_{{i}}}}}} \right)^ + },\\
{{P'}_1} + {{P'}_2} = {P_1} + {P_2} - \left( {{{P''}_1} + {{P''}_2}} \right).
\end{array}
\end{equation}

From \eqref{max-solution-fading}, we observe that when ${p_{{ni}}}\left( {{P_I} - 1} \right) > {P_1} + {P_2}$, the sum-rate optimal strategy is to avoid transmission when PU is active, and this is identical to the three-switch model. Also as expected, it suggests that when $P_I$ is large or $p_i$ is small, performance gain through decoding in the presence of PU would be very limited. Numerical results in Section V validate our analysis, where marginal performance improvement is observed even when $P_I$ is not large.

We continue to include channel fading into consideration, and explore the following model
\begin{equation}
Y = \left\{ {\begin{array}{*{20}{c}}
{{S_{{T_1}}}{\sqrt{H_1}}{X_1} + {S_{{T_2}}}{\sqrt{H_2}}{X_2} + Z,\quad {S_R} = 1},\\
{{S_{{T_1}}}{\sqrt{H_1}}{X_1} + {S_{{T_2}}}{\sqrt{H_2}}{X_2} + I,\quad {S_R} = 0},
\end{array}} \right.
\end{equation}
where $\sqrt{H_1}$ and $\sqrt{H_2}$ are the channel gains between the cognitive transmitters and receiver.

By letting ${\sqrt{h_1}} = \frac{{{S_{{T_1}}}{\sqrt{H_1}}}}{{{S_R} - \left( {{S_R} - 1} \right)\sqrt {{P_I}} }}$ and ${\sqrt{h_2}} = \frac{{{S_{{T_2}}}{\sqrt{H_2}}}}{{{S_R} - \left( {{S_R} - 1} \right)\sqrt {{P_I}} }}$, the channel model can be normalized as $
Y = {\sqrt{h_1}}{X_1} + {\sqrt{h_2}}{X_2} + Z$.

Define ${\bf{h}} = \left[ {{h_1},{h_2}} \right]$ as the normalized
channel gain vector and consider all fading possibilities, whose cdf and pdf are denoted by $F_i(h_i)$ and $f_i(h_i)$, respectively. The
achievable rate region of the cognitive MAC fading channel \cite{MAC-fading-D.Tse} can be expressed as:
\begin{equation}
\begin{aligned}
&{C^{{\rm{fading}}}}\left( {{P_1},{P_2}} \right) = \\
&\bigcup\limits_{\textbf{P} \in {\cal F}} {\left\{ {\begin{array}{*{20}{c}}
{{R_1} \le {E_{\bf{h}}}\left[ {\log \left( {1 + {h_1}{P_1}\left( {\bf{h}} \right)} \right)} \right]}\\
{{R_2} \le {E_{\bf{h}}}\left[ {\log \left( {1 + {h_2}{P_2}\left( {\bf{h}} \right)} \right)} \right]}\\
{{R_1} + {R_2} \le {E_{\bf{h}}}\left[ {\log \left( {1 + \sum\limits_{{i}} {{h_i}{P_i}\left( {\bf{h}} \right)} } \right)} \right]}
\end{array}} \right\}} ,
\end{aligned}
\end{equation}
{\flushleft{where}} ${\cal F} \equiv \left\{ {\textbf{P}:{E_{\bf{h}}}\left[
{{P_i}({\bf{h}})} \right] \le {P_i},  \forall i \in 1,2} \right\}$
is the set of all possible power allocation schemes within the power constraints.

Now, we discuss the optimal rate/power allocation under
fading and interference when the PU states and channel gains are
known to both the transmitters and receiver. According to
\cite{MAC-fading-D.Tse}, when global state information is available, the maximum sum rate ${R_1} + {R_2}$ can be optimized over all possible power allocation schemes. The solution has the form of:
\begin{align}\label{power-allocation-D.Tse}
P_1^*\left( {{\bf{h}},{\bf{\lambda }}} \right) = \left\{ {\begin{array}{*{20}{c}}
{{{\left( {\frac{1}{{2{\lambda _1}}} - \frac{1}{{{h_1}}}} \right)}^ + ,}}&{{h_1} \ge \frac{{{\lambda _1}}}{{{\lambda _2}}}{h_2}},\\
0,&{else,}
\end{array}} \right.\nonumber \\
P_2^*\left( {{\bf{h}},{\bf{\lambda }}} \right) = \left\{ {\begin{array}{*{20}{c}}
{{{\left( {\frac{1}{{2{\lambda _2}}} - \frac{1}{{{h_2}}}} \right)}^ + ,}}&{{h_2} \ge \frac{{{\lambda _2}}}{{{\lambda _1}}}{h_1}},\\
0,&{else,}
\end{array}} \right.
\end{align}
{\flushleft{where}} the constant vector ${\bf{\lambda }} = \left[ {{\lambda
_1},{\lambda _2}} \right]$ is determined by the following average
power constraints:
\begin{align*}
\int_0^\infty  {{{\left( {\frac{1}{{2{\lambda _1}}} - \frac{1}{h}} \right)}^ + }F_1\left( {\frac{{{\lambda _1}}}{{{\lambda _2}}}h} \right)f_1\left( h \right)} dh \le {P_1},\\
\int_0^\infty  {{{\left( {\frac{1}{{2{\lambda _2}}} - \frac{1}{h}} \right)}^ + }F_2\left( {\frac{{{\lambda _2}}}{{{\lambda _1}}}h} \right)f_2\left( h \right)} dh \le {P_2}.
\end{align*}

\emph{Remarks:} From \eqref{power-allocation-D.Tse}, we can see that for the on/off cognitive MAC channel with PU interference and fading, the optimal rate/power allocation is given by the generalized time-domain water-filling, which is performed on the generalized parameters ${h_1}$ and ${h_2}$, taking into account the factors of the PU states and PU interference.

When ${h_1}$ and ${h_2}$ are identically distributed, ${\lambda _1} = {\lambda _2}$ by symmetry.
An observation from \eqref{power-allocation-D.Tse} is
that when the channel gain of one SU transmitter is worse than
another, the transmission will be turned off to save power. In other
words, the two SU transmitters will transmit simultaneously with
equal power only when their channel gains are the same. Therefore,
the cognitive MAC channel with fading and interference can be
further simplified into the following model:
\begin{equation}\label{FadingModelSimplified}
Y = \left\{ {\begin{array}{*{20}{c}}
{S{'_{{T_1}}}{X_1} + S{'_{{T_2}}}{X_2} + Z',\quad {S_R} = 1},\\
{S{'_{{T_1}}}{X_1} + S{'_{{T_2}}}{X_2} + I',\quad {S_R} = 0},
\end{array}} \right.
\end{equation}
where {${S'_{{T_1}}} = \left\{ {\begin{array}{*{20}{c}}
{1,}&{{S_{{T_1}}} = 1\;{\rm{and}}\;{H_1} \ge {H_2}},\\
{0,}&{{S_{{T_1}}} = 0\;{\rm{or}}\;{H_1} < {H_2}},
\end{array}} \right.$} and {${S'_{{T_2}}} = \left\{ {\begin{array}{*{20}{c}}
{1,}&{{S_{{T_2}}} = 1\;{\rm{and}}\;{H_1} \le {H_2},}\\
{0,}&{{S_{{T_2}}} = 0\;{\rm{or}}\;{H_1} > {H_2}},
\end{array}} \right.$} are the generalized binary transmitter side information, $Z' = \frac{Z}{{\max \left\{ {\sqrt {{H_1}} ,\sqrt {{H_2}} } \right\}}}$ and $I' = \frac{I}{{\max \left\{ {\sqrt {{H_1}} ,\sqrt {{H_2}} } \right\}}}$ are the generalized noise and interference. Note that \eqref{FadingModelSimplified} has the same form as \eqref{InterferenceModel}, so similar methods can be employed to obtain the maximum sum rate.

\subsection{Extension to the m-user Case}
Throughout this paper, we focus on the three-switch channel where we only consider two transmitters. However, the results and analysis
can be generalized to the $m$-user case. The extension is omitted here due to space limitation. The reader is referred to
\cite{TechReport-CogMAC} for the details.

\section{Numerical Results}\label{SecSim}

\subsection{Outer and Inner Bounds}
To plot the outer and inner bounds, we travel through all possible
power pairs which satisfy the power constraints, and compute the corresponding
rate pairs. As for the inner bound, we calculate the gap between \emph{outer bound 2} and
\emph{inner bound} and subtract it from \emph{outer bound 2}.

\begin{figure}[t] \centering
\includegraphics[width=0.4\textwidth]{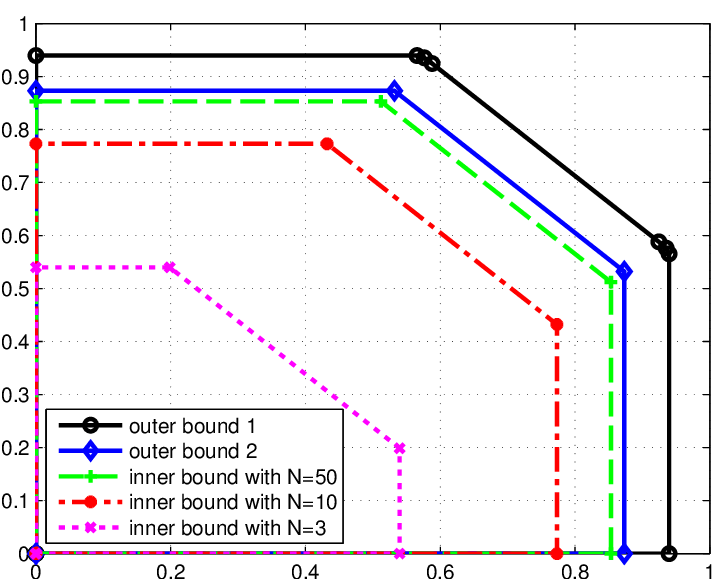}
\caption {Capacity region bounds when $\mu=0.1, P=1, \rho=0$} \label{bound1}
\end{figure}

\begin{figure}[t] \centering
\includegraphics[width=0.4\textwidth]{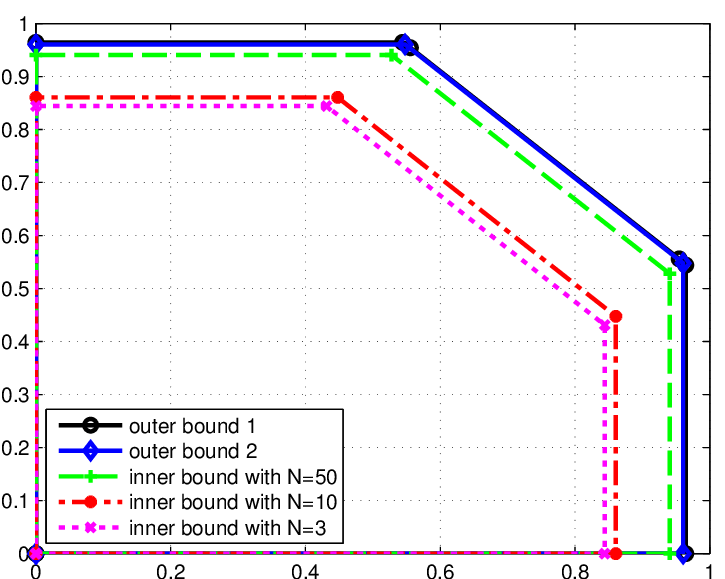}
\caption {Capacity region bounds when $\mu=0.1, P=1, \rho=0.9$}
\label{bound2}
\end{figure}

\begin{figure}[t] \centering
\includegraphics[width=0.4\textwidth]{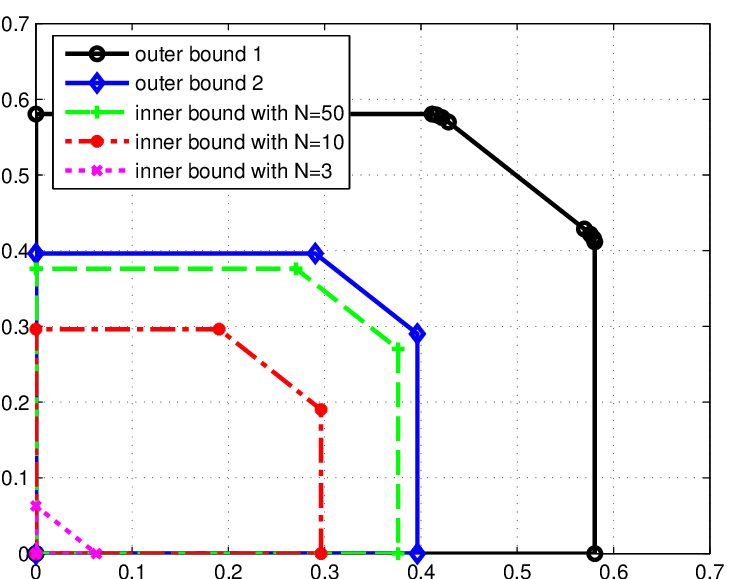}
\caption {Capacity region bounds when $\mu=0.5, P=1, \rho=0$} \label{bound3}
\end{figure}

\begin{figure}[t] \centering
\includegraphics[width=0.4\textwidth]{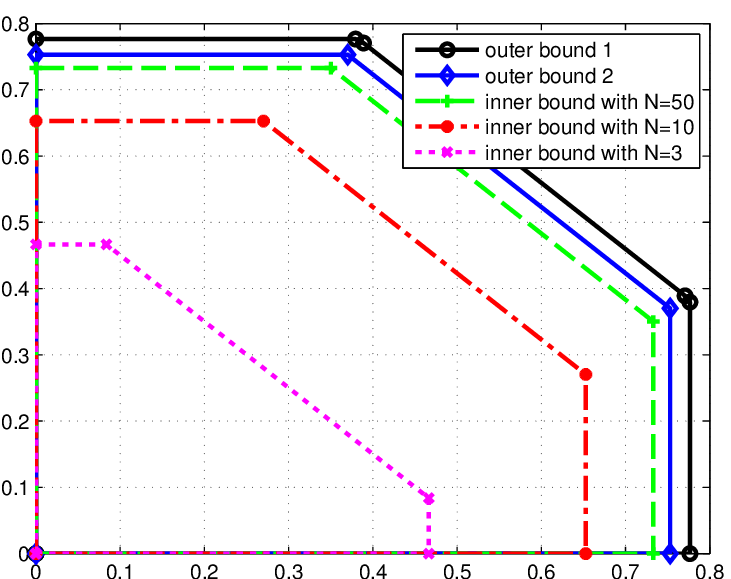}
\caption {Capacity region bounds when $\mu=0.5, P=1, \rho=0.9$}
\label{bound4}
\end{figure}

\begin{figure}[t] \centering
\includegraphics[width=0.4\textwidth]{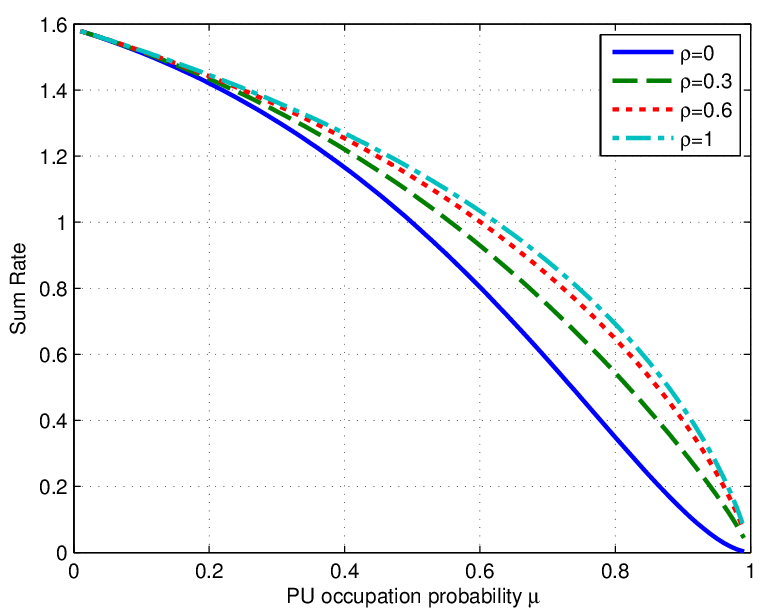}
\caption {Effect of PU activities on sum rate} \label{Sumrate_mu}
\end{figure}

\begin{figure}[t] \centering
\includegraphics[width=0.4\textwidth]{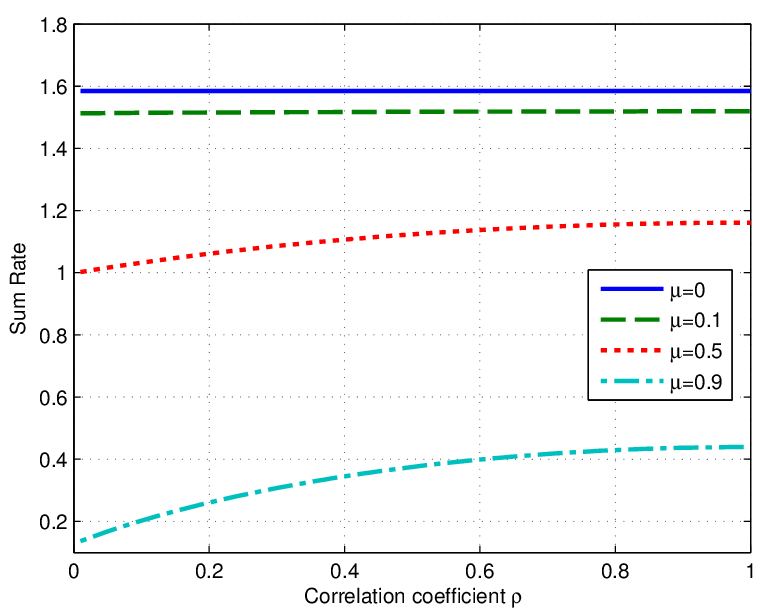}
\caption {Effect of side information's correlation on sum rate}
\label{Sumrate_rho}
\end{figure}

Fig.~\ref{bound1} to Fig.~\ref{bound4} show the \emph{outer bound 1,
2} and the \emph{inner bound} under different PU occupation rates
$\mu$ and PU states correlation coefficients $\rho$. The parameters
are $(\mu=0.1, \rho=0)$, $(\mu=0.1, \rho=0.9)$, $(\mu=0.5, \rho=0)$
and $(\mu=0.5, \rho=0.9)$ for the four figures, respectively. All
four cases are under the power constraints of $P_1 \le 1$ and $P_2
\le 1$. Based on these parameters, we calculate $p_a$ to $p_f$
according to \eqref{probabilities}, and obtain the bounds.
We also plot the sum rate of \emph{outer bound 1} with
respect to PU occupation rate $\mu$ and correlation coefficient
$\rho$ on Fig.~\ref{Sumrate_mu} to Fig.~\ref{Sumrate_rho}, in which
the sum rate corresponds to the corner points in Fig.~\ref{bound1}
to Fig.~\ref{bound4}.

The insights obtained from these results are summarized below.
\begin{enumerate}
\item{Correlation coefficient $\rho$:} By comparing Fig.~\ref{bound1} with
Fig.~\ref{bound2}, and Fig.~\ref{bound3} with Fig.~\ref{bound4}, we
find gaps between the outer and inner bounds become closer as $\rho$
grows. On the one hand, \emph{outer bound 2} approaches \emph{outer bound 1} as $\rho$ grows, as the transmitters in \emph{Case 2}
gradually have global side information, and \emph{Case 2} evolves to
\emph{Case 1}. On the other hand, \emph{inner bound}
approaches \emph{outer bound 2} because $ H\left( {S_{T_i }\left( t \right) \left|
{S_R\left( t \right) } \right.} \right)$ decreases as $\rho$ grows.
In addition, Fig.~\ref{Sumrate_rho} also demonstrates that the
maximal sum rate increases with $\rho$, which indicates more
correlation improves the overall performance.

\item{PU occupation rate $\mu$:} By comparing Fig.~\ref{bound1} with
Fig.~\ref{bound3}, and Fig.~\ref{bound2} with Fig.~\ref{bound4}, we
observe that as $\mu$ decreases, the overall
system throughput grows. This is further verified in Fig.~\ref{Sumrate_mu}. When $\mu = 0$, the PUs keep silent, and the
cognitive MAC rate region evolves to the traditional MAC rate region. When $\mu = 1$, the spectrum is always occupied by PUs, the
cognitive MAC rate degrades to zero.

\item{State changing rate:} Fig.~\ref{bound1} to Fig.~\ref{bound4} show that, the
inner bound are closer to the outer bounds when the $N$ increases.
This indicates that, in the case where the PU states change
slowly, the \emph{inner bound} approaches \emph{outer bound 2}.

\item{Transmitter Side information:} Compared with \emph{Case 1}, the transmitters in
\emph{Case 2} lack global side information. Hence, the big gap
between \emph{outer bound 1} and \emph{2} implies the importance of the transmitter side information. Moreover, this gap is larger in Fig.~\ref{bound3} than in Fig.~\ref{bound1}, which indicates that the transmitter side information is more valuable in busy channels.

\end{enumerate}

%\subsection{Optimal Rate/Power Allocation}
%
%We adopt the same parameters as in Section V.A, and plot the
%sum rate $R_1+R_2$ according to \emph{Corollary 1}. The power
%allocation results are illustrated in Fig.~\ref{1power} and
%Fig.~\ref{2power}. The PU occupation rates for both figures are
%$\mu=0.1$, while the correlation coefficients $\rho$ are $0$ and
%$0.5$, respectively.
%
%Again we assume the average power constraints are $P_1 \le 1$ and $P_2 \le
%1$, and the top curve corresponds to the case when
%maximal power is consumed, i.e. $ p_a P_1^a  + p_c P_1^c  = 1 $
%and $p_b P_2^b + p_c P_2^c  = 1$. Since we choose the same values
%of $\mu$ and $\rho$ for both transmitters and the receiver, the
%power allocation results are symmetric between the two transmitters,
%i.e. $ P_1^a  = P_2^b ,P_1^c = P_2^c$. As shown in Fig.~\ref{1power}
%and Fig.~\ref{2power}, different power allocations generate
%different sum rates. Since \eqref{optimization} is a convex
%optimization problem, we can always find an optimal power allocation
%scheme which maximize the sum rate. Note that here the individual powers may exceed $1$ because our constraints are imposed on the average power, therefore the peak powers can be higher.

\subsection{Optimized Sum Rate in Interference Model}
\begin{figure}[t] \centering
\includegraphics[width=0.4\textwidth]{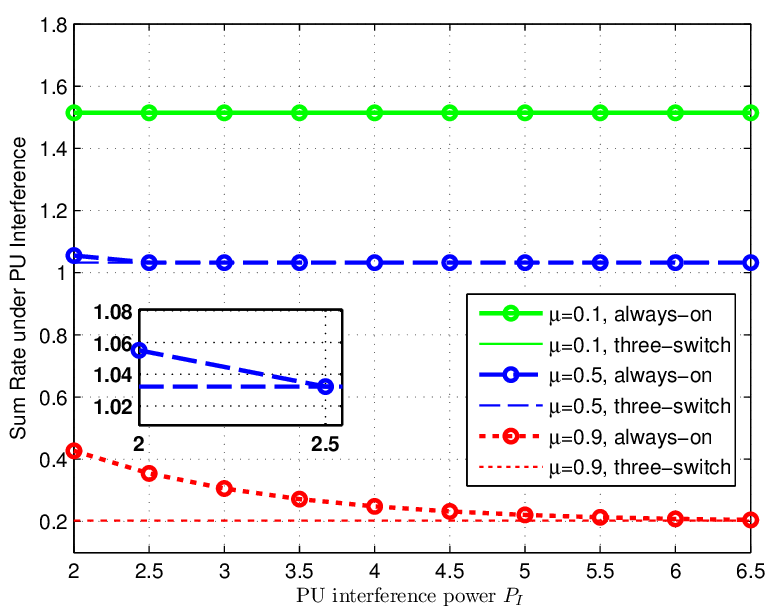}
\caption {Sum rate in interference model v.s. $P_I$}
\label{Sumrate_P_I_fading}
\end{figure}

\begin{figure}[t] \centering
\includegraphics[width=0.4\textwidth]{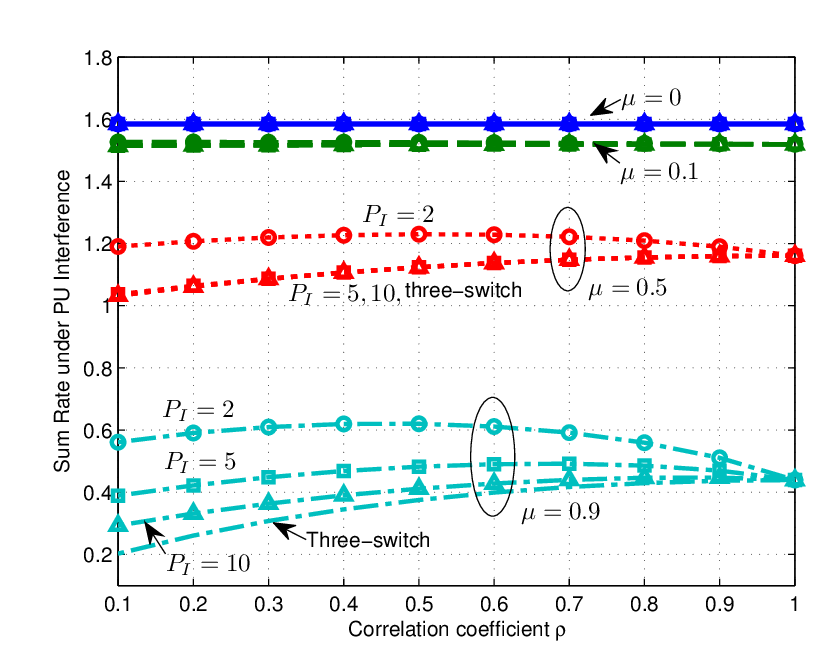}
\caption {Sum rate in interference model v.s. $\rho$}
\label{Sumrate_rho_fading}
\end{figure}

We now provide numerical results for the cognitive MAC channel when the receiver is exposed to PU interference (c.f. \eqref{InterferenceModel}), which is discussed in Section VI.E. The joint probabilities are shown in Table I. In addition, we assume that the power constraints are $P_1 \le 1$ and $P_2 \le 1$. The maximal sum rates with respect to the PU interference power $P_I$, PU occupation rate $\mu$ and the correlation coefficient $\rho$ are plotted in Fig.~\ref{Sumrate_P_I_fading} and Fig.~\ref{Sumrate_rho_fading}, which are computed according to \eqref{probabilities}, \eqref{max-sumrate-fading}, and \eqref{max-solution-fading}.

Suppose that the receiver employs an energy detector with a threshold 2W, through which it determines if a PU is present. Fig.~\ref{Sumrate_P_I_fading} shows that the advantage of keeping the receiver always on (and decoding) over the original three-switch model is significant only for small $P_I$ values and when $\mu$ is large (i.e., the PU is active). Of course, the performance of model \eqref{InterferenceModel} is no worse than that of \eqref{cogMAC} concerning the optimal sum rate, due to the optimal rate/power allocation, which essentially turns off transmission in the face of severe interference.

In Fig.~\ref{Sumrate_rho_fading}, $P_I$ is selected as $2$W, $5$W and $10$W and the following two insights on the additional benefit brought by allowing the secondary receiver to
remain receiving even in the presence of PU interference are further confirmed.
\begin{enumerate}
\item{It is rewarding to turn on the receiver's switch only when the PUs are very active: }when $\mu$ is large, the additional time period that the receiver can decode is longer, which results in more benefit. As seen, the additional benefit is obvious only when $\mu>0.5$.
\item{More opportunities can be exploited under lower correlation among PU states: }this is due to the fact that when $\rho$ is large, the transmitters are likely to keep silent when the receiver is under PU interference.
\end{enumerate}

\section{Conclusion}
In this paper, the cognitive MAC channel is modeled as a three-switch channel, and the achievable rate regions are obtained when viewing PU activities as causal/non-causal on/off side information. The closed form outer and inner bounds are derived, which are shown to be tight under some special cases. A rate allocation scheme is also proposed to maximize the sum rate with global side information, and the effect of correlation in side information and PU activities is analyzed. The extension to the fading scenario and general $m$-user case is discussed. The numerical results show the importance of transmitter side information in enhancing the capacity and the effectiveness of our rate allocation scheme.

%\section{Acknowledgement}
%
%The authors gratefully thank Yu Zhang and Rui Yin of Zhejiang University for their valuable advices during multiple discussions and Xiao Cai also from Zhejiang University for his contribution to the simulation part.

%\vspace{0.25in}

\bibliographystyle{unsrt}

\begin{thebibliography}{100}

\bibitem{NSFreport}
M. A. McHenry, ``NSF spectrum occupancy measurements project summary," \emph{Shared Spectrum Company Report}, Aug. 2005.

\bibitem{FCCorder}
``In the matter of unlicensed operation in the TV broadcast bands:
second report and order and memorandum opinion and order,"
Federal Communications Commision, Tech. Rep. 08-260, Nov. 2008.
[Online]. Available: http://hraunfoss.fcc.gov/edocs\_public/attachmatch/
FCC-08-260A1.pdf.

%\bibitem{CogSeNet}
%H. G. Goh, K. H. Kwong, C. Shen, C. Michie, and I. Andonovic,
%``CogSeNet: a concept of cognitive wireless sensor network," in
%\emph{Proc. IEEE CCNC}, pp. 1-2, Jan. 2010.

\bibitem{CRSN}
O. Akan, O. Karli, O. Ergul, and M. Haardt, ``Cognitive radio sensor networks," \emph{IEEE Network}, vol. 23, no. 4, pp. 34-40 July 2009.
%
%\bibitem{CWSN}
%A. S. Zahmati, S. Hussain, X. Fernando, and A. Grami, ``Cognitive
%wireless sensor networks: emerging topics and recent challenges," in \emph{Proc. IEEE TIC-STH}, pp. 593-596 Sept. 2009.
%
%\bibitem{CWSNsurvey}
%G. Vijay, E. Bdira, and M. Ibnkahla, ``Cognitive approaches in
%wireless sensor networks: a survey," in \emph{Proc. QBSC}, pp. 177-180, May 2010.
%
%\bibitem{CRWSN}
%K. L. A. Yau, P. Komisarczuk, P. D. Teal, ``Cognitive radio-based
%wireless sensor networks: conceptual design and open issues," in
%\emph{Proc. LCN}, pp. 955-962, Oct. 2009.

%\bibitem{SP-CR}
%J. Ma, G. Y. Li, B. H. Juang, ``Signal processing in cognitive
%radio," \emph{Proceedings of the IEEE}, vol.97, no.5, pp.805-823,
%May 2009.

\bibitem{802.22}
IEEE Std 802.22-2011(TM), Standard for Wireless Regional Area Networks-Part 22: Cognitive Wireless RAN Medium Access Control (MAC) and Physical Layer (PHY) specifications: policies and procedures for operation in the TV bands, July 2011.

\bibitem{Cluster-CRSN}
H. Zhang, Z. Zhang, H. Dai, R. Yin and X. Chen, ``Distributed
spectrum-aware clustering in cognitive radio sensor networks, in
\emph{Proc. IEEE GlobeCom}, Dec. 2011.

\bibitem{Zhang-2011ICC-JSCS}
H. Zhang, Z. Zhang, X. Chen and R. Yin, ``Energy Efficient Joint Source and Channel Sensing in Cognitive Radio Sensor Networks," in \emph{Proc. IEEE ICC}, June 2011.

\bibitem{Zhang-2011INFOCOM-CS}
H. Zhang, Z. Zhang, Y. Chau, ``Distributed compressed wideband sensing in Cognitive Radio Sensor Networks," in \emph{Proc. IEEE INFOCOM WKSHPS}, April 2011.

\bibitem{Gridlock-CR-IT}
A. Goldsmith, S. A. Jafar, I. Maric and S. Srinivasa, ``Breaking spectrum gridlock with cognitive radios: an information theoretic perspective"
\emph{Proceedings of the IEEE}, vol. 97, no. 5, pp. 894-914, May 2009.

\bibitem{Underlay}
A. Ghasemi and E. S. Sousa, ``Capacity of fading channels under
spectrum-sharing constraints," in \emph{Proc. IEEE ICC}, June 2006.

\bibitem{Overlay1}
A. Jovicic and P. Viswanath, ``Cognitive radio: an
information-theoretic perspective," \emph{IEEE Trans. Inf.
Theory}, vol. 55, no. 9, pp. 3945-3958, Sept. 2009.

\bibitem{Overlay2}
N. Devroye, P. Mitran and V. Tarokh, ``Achievable rates in cognitive radio channels," \emph{IEEE Trans. Inf. Theory}, vol. 52, no. 5,
pp. 1813-1827, May 2006.

\bibitem{Overlay3-ICDMS}
W. Wu, S. Vishwanath, and A. Arapostathis, ``Capacity of a class of
cognitive radio channels: interference channels with degraded
message sets," \emph{IEEE Trans. Inf. Theory}, vol. 53, no. 11,
pp. 4391-4399, Nov. 2007.

\bibitem{Overlay4-ICDMS}
J. Jiang and Y. Xin, ``On the achievable rate regions for
interference channels with degraded message sets," \emph{IEEE Trans. Inf. Theory}, vol. 54, no. 10, pp. 4707-4712, Oct. 2008.


\bibitem{fading-C-MAC}
R. Zhang, S. Cui, and Y. C. Liang, ``On ergodic sum capacity of fading cognitive multiple-access and broadcast channels," \emph{IEEE Trans. Inf. Theory}, vol. 55, no. 11, pp. 5161-5178, Nov. 2009.

\bibitem{MID}
R. Zhang and Y. C. Liang, ``Investigation on multiuser diversity in spectrum sharing based cognitive radio networks," \emph{IEEE Commun. Letters}, vol. 14, no.2, pp. 133-135, Feb. 2010.

\bibitem{NetworkCapacity}
Piyush Gupta and P. R. Kumar, ``The capacity of wireless networks,"
\emph{IEEE Trans. Inf. Theory}, vol. 42, no. 2, pp. 388-404,
March 2000.

\bibitem{NetworkCapacityCR}
S. Jeon, N. Devroye, M. Vu, S. Chung, and V. Tarokh, ``Cognitive
networks achieve throughput scaling of a homogeneous network," to
appear in \emph{IEEE Trans. Inf. Theory}, vol.57, no.8, pp.5103-5115, Aug. 2011.

\bibitem{NetworkCapacityCR2}
C. Yin, L. Gao, and S. Cui, ``Scaling laws for overlaid wireless
networks: A cognitive radio network vs. a primary network,"
\emph{IEEE Trans. Networking}, vol. 18, no. 4, pp. 1317-1329, Aug.
2010.

\bibitem{Highway}
M. Franceschetti, O. Dousse, D. Tse, and P. Thiran, ``Closing the
gap in the capacity of wireless networks via percolation theory,"
\emph{IEEE Trans. Inf. Theory}, vol. 53, no. 3, pp. 1009-1018,
Mar. 2007.

\bibitem{NetworkCapacityCR_CLi}
C. Li and H. Dai, ``On the throughput scaling of cognitive radio ad
hoc networks," \emph{IEEE INFOCOM}, mini conference, Shanghai,
China, Apr. 2011.

\bibitem{Sun-Achievable-RelayCRN}
L. Sun and W. Wang, ``On Study of Achievable Capacity with Hybrid Relay in Cognitive Radio Networks, in \emph{Proc. IEEE GlobeCom}, Nov. 2009.


\bibitem{MAC-side}
S. A. Jafar, ``Capacity with causal and noncausal side information:
a unified view," \emph{IEEE Trans. Inf. Theory},
vol. 52, no. 12, pp. 5468-5474, Dec. 2006.

\bibitem{Two-switch}
S. A. Jafar and S. Srinivasa, ``Capacity limits of cognitive radio
with distributed and dynamic spectral activity," \emph{IEEE JSAC},
vol. 25, no. 3, pp. 529-537, April 2007.

\bibitem{SpectrumSensing}
J. Ma, G. Y. Li, and B. H. Juang, ``Signal processing in cognitive
radio," \emph{Proceedings of the IEEE}, vol. 97, no. 5, pp. 805-823, May 2009.

\bibitem{SideInfom}
C. E. Shannon, ``Channels with side information at the transmitter," \emph{IBM Journal of Research and Development}, vol. 2, pp. 289-293, 1958.

\bibitem{determ}
T. M. Cover and M. Chiang, ``Duality between channel capacity and
rate distortion with two-sided state information," \emph{IEEE
Trans. Inf. Theory}, vol. 48, no. 6, pp. 1629-1638, June 2002.

\bibitem{MAC-fading-D.Tse}
D. Tse and S. Hanly, ``Multi-access fading channels: part I: polymatroid structure, optimal resource allocation and throughput capacities", \emph{IEEE Trans. Inf. Theory}, vol. 44, no. 7, pp. 2796-2815, Nov. 1998.

\bibitem{TechReport-CogMAC}
H. Zhang, Z. Zhang and H. Dai, ``On the Capacity Region of Cognitive Multiple Access
over White Space Channels," Technical report, NC State University,
Department of Electrical Engineering, 2012.

%\bibitem{JCSC-ICC}
%H. Zhang, Z. Zhang, X. Chen and R. Yin, ``Energy Efficient Joint
%Source and Channel Sensing in Cognitive Radio Sensor Networks,"
%Accepted by \emph{Proc. IEEE ICC 2011}, June 2011.


%\bibitem{Distributed-Source-Coding}
%Z. Xiong; A. D. Liveris, S. Cheng, ``Distributed source coding for
%sensor networks," \emph{IEEE Signal Processing Magzine}, vol.21,
%no.5, pp.80-94, Sept. 2004.
%
%\bibitem{MDC}
%V. K. Goyal, ``Multiple description coding: compression meets the
%network," \emph{IEEE Signal Processing Magzine}, vol.18, no.5,
%pp.74-93, Sept. 2001.
%
%\bibitem{CEO}
%T. Berger, Z. Zhang, H. Viswanathan, ``The CEO problem
%[multiterminal source coding]," \emph{IEEE Transactions on
%Information Theory}, vol.42, no.3, pp.887-902, May 1996.




\end{thebibliography}

% that's all folks

\begin{biography}[{\includegraphics[width=1.0\textwidth] {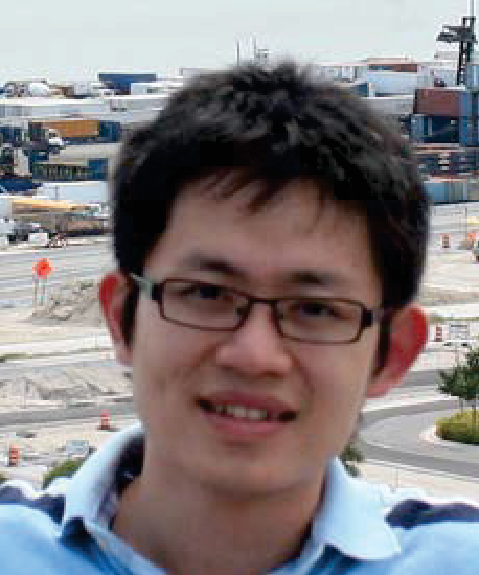}}]{Huazi Zhang}
received the B.S. degree in communication engineering from Zhejiang University, Hangzhou, China, in 2008. He then began to pursue his Ph.D degree in the Department of Information Science and Electronic Engineering, Zhejiang University. He is currently a visiting Ph.D student in the Department of Electrical and Computer Engineering, NC State University, Raleigh. His research interests include multiuser information theory, cognitive radio and networked information processing in mobile ad hoc networks and sensor networks.
\end{biography}

\begin{biography}[{\includegraphics[width=1.0\textwidth] {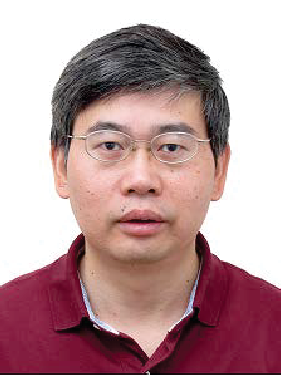}}]{Zhaoyang Zhang}
[M'02] received his B.Sc. and Ph.D. degrees from Zhejiang University, Hangzhou, China,
in 1994 and 1998, respectively. He is currently a professor with the
Department of Information Science and Electronic Engineering, Zhejiang University. He was selected into the Supporting Program for New
Century Excellent Talents in University (NCET) by the Ministry of
Education, China, in 2009. His current research interests are
mainly focused on network information theory and
advanced coding theory, network signal processing,
cognitive radio networks and cooperative relay networks, etc., as well as their
applications in next generation wireless communications. He has authored
or co-authored more than 150 refereed journal and conference papers in the above areas. He is currently
serving as an Associate Editor for the \emph{International
Journal of Communication Systems}.
\end{biography}

\begin{biography}[{\includegraphics[width=1.0\textwidth] {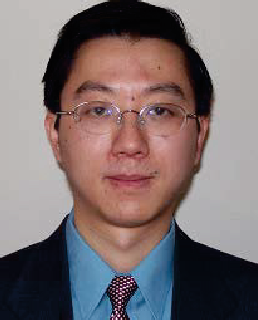}}]{Huaiyu Dai}
[M'03, SM'09] received the B.E. and M.S. degrees in electrical engineering from Tsinghua University, Beijing, China, in 1996 and 1998, respectively, and the Ph.D. degree in electrical engineering from Princeton University, Princeton, NJ in 2002.

He was with Bell Labs, Lucent Technologies, Holmdel, NJ, during summer 2000, and with AT\&T Labs-Research, Middletown, NJ, during summer 2001. Currently he is an Associate Professor of Electrical and Computer Engineering at NC State University, Raleigh. His research interests are in the general areas of communication systems and networks, advanced signal processing for digital communications, and communication theory and information theory. His current research focuses on networked information processing and crosslayer design in wireless networks, cognitive radio networks, wireless security, and associated information-theoretic and computation-theoretic analysis.

He has served as editor of IEEE Transactions on Communications, Signal Processing, and Wireless Communications. He co-edited two special issues for EURASIP journals on distributed signal processing techniques for wireless sensor networks, and on multiuser information theory and related applications, respectively.
\end{biography}

%\begin{figure}[t] \centering
%\includegraphics[width=0.6\textwidth]{1power.eps}
%\caption {Optimal power allocation on outer bounds 1, $\mu=0.1,
%\rho=0$} \label{1power}
%\end{figure}
%
%\begin{figure}[t] \centering
%\includegraphics[width=0.6\textwidth]{2power.eps}
%\caption {Optimal power allocation on outer bounds 1, $\mu=0.1,
%\rho=0.5$} \label{2power}
%\end{figure}

\end{document}